\documentclass[preprint,tightenlines,aps,nofootinbib,showpacs]{revtex4}

\usepackage{amsmath,amssymb,array}
\usepackage{graphicx}
\usepackage[dvips]{color}

\newcommand{\notE}{\hbox{{$E$}\kern-.60em\hbox{/}}}
\newcommand{\notp}{\hbox{{$p$}\kern-.43em\hbox{/}}}
\def\D0{\mbox{D\O}}
\newcolumntype{L}[1]{>{\raggedright\let\newline\\\arraybackslash\hspace{0pt}}m{#1}}
\newcolumntype{C}[1]{>{\centering\let\newline\\\arraybackslash\hspace{0pt}}m{#1}}
\newcolumntype{R}[1]{>{\raggedleft\let\newline\\\arraybackslash\hspace{0pt}}m{#1}}
\preprint{\font\fortssbx=cmssbx10 scaled \magstep2
\hbox to \hsize{
\hskip1.2in 
\hbox{\fortssbx The University of Oklahoma}
\hskip0.2in $\vcenter{
                      \hbox{\bf arXiv: [hep-ph]}
                      \hbox{\bf OU-HEP-181225}
                      \hbox{May 2019}}$ }
}

\begin{document}


\title{\vspace*{0.7in}
Flavor Changing Heavy Higgs Interactions with Leptons at Hadron Colliders}

\author{
Wei-Shu Hou$^{a}$\footnote{E-mail address: wshou@phys.ntu.edu.tw},
Rishabh Jain$^{b}$\footnote{E-mail address: Rishabh.Jain@ou.edu},
Chung Kao$^{b}$\footnote{E-mail address: Chung.Kao@ou.edu},
Masaya Kohda$^{a}$\footnote{E-mail address: mkohda@hep1.phys.ntu.edu.tw},
Brent McCoy$^{b}$\footnote{E-mail address: mccoy@physics.ou.edu}
Amarjit Soni$^{c}$\footnote{E-mail address: adlersoni@gmail.com}
}

\affiliation{
$^a$Department of Physics, National Taiwan University,
Taipei 10617, Taiwan, ROC \\
$^b$Homer L. Dodge Department of Physics and Astronomy,
University of Oklahoma, Norman, OK 73019, USA \\
$^c$Physics Department, Brookhaven National Laboratory, Upton, NY
11973, USA}

\date{\today}

\bigskip

\begin{abstract}

In a general two Higgs doublet model, we study
flavor changing neutral Higgs (FCNH) decays into leptons at hadron colliders,
$pp \to \phi^0 \to \tau^\mp\mu^\pm +X$, where $\phi^0$ could be
a CP-even scalar ($h^0$, $H^0$)
or a CP-odd pseudoscalar ($A^0$).
The light Higgs boson $h^0$ is found to resemble
closely the Standard Model Higgs boson at the Large Hadron Collider.
In the alignment limit of $\cos(\beta-\alpha) \cong 0$ for $h^0$--$H^0$ mixing,
FCNH couplings of $h^0$ are naturally suppressed,
but such couplings of the heavier $H^0, A^0$
are sustained by $\sin(\beta-\alpha) \simeq 1$.
We evaluate physics backgrounds from dominant processes
with realistic acceptance cuts and tagging efficiencies.
We find promising results for
$\sqrt{s} = 14$ TeV,
which we extend further to $\sqrt{s} = 27$ TeV and 100 TeV
future pp colliders.
\end{abstract}

%

\maketitle

\newpage

\section{Introduction}

Recent 13 TeV studies at the LHC by ATLAS and CMS experiments
confirm that the properties of the 125 GeV
Higgs boson are in good agreement with the expectations from the
Standard Model (SM) Higgs boson~\cite{Sirunyan:2018koj,ATLAS:2019slw}. 
This is in sharp contrast to the persistent signs of significant
deviation from SM in the flavor sector.
The 3--4$\sigma$ hints of lepton universality violation,
in simple tree-level semi-leptonic B decays  as well as
in flavor-changing loop processes, have been  very much
in the news~\cite{flavor_anoms_2018}.
Moreover, for over a decade now the muon anomalous magnetic moment measurement
at BNL~\cite{Bennett:2006fi} also seem to show about 3.5$\sigma$ deviation
from SM.
While so far none of these constitutes compelling evidence against
the SM,
but even if just one of them pans out, it would be physics beyond SM.
In particular, it is important to recall that lepton universality is
purely an accidental symmetry of SM.
These underpinnings prompt us to question lepton universality and lepton
flavor violation in the Higgs sector itself.

Our investigation was motivated by the experimental 2$\sigma$ hint
for $h^0 \to \tau \mu$ from CMS~\cite{Khachatryan:2015kon}.
While it has subsequently disappeared~\cite{Sirunyan:2017xzt},
it in fact motivates further the search for $H^0,\, A^0 \to \tau\mu$,
involving {\it heavy} exotic Higgs bosons, as we shall explain.
Note in particular that, in face of the current semileptonic anomalies
in B decays, a general two Higgs doublet model (g2HDM)
had been invoked~\cite{Crivellin:2012ye}
over the disfavored conventional Type II of two Higgs doublet models (2HDM-II).
While the situation with the anomalies are as yet inconclusive,
we adopt the g2HDM set up in this work, i.e. without the usual
$Z_2$ symmetry to forbid flavor changing neutral Higgs (FCNH) couplings.
Another mechanism may be at work instead of $Z_2$ or
Natural Flavor Conservation~\cite{Glashow:1976nt}:
alignment~\cite{Hou:2017hiw}.
Removing interactions of the extra scalars with vector boson pairs
($H^0 WW$ and $H^0 ZZ$), other than the SM-Higgs,
is known as the alignment limit~\cite{
Gunion:2002zf,Carena:2013ooa,Dev:2014yca}.
Influenced by the LHC results on the 125 GeV
boson~\cite{Sirunyan:2018koj,ATLAS:2019slw}, 
we will assume that one must work close to this limit.

We seek the discovery of the leptonic flavor changing decay,
specifically $\phi^0 \to \tau \mu$, where $\phi^0 = h^0,H^0,A^0$.
In SM, $h^0 \to \tau \mu$ is highly suppressed
at loop level by the extremely tiny neutrino masses,
but in g2HDM without any $Z_2$ symmetry, this decay is
in principle possible at {\it tree} level.
We adopt the following interaction
Lagrangian~\cite{Davidson:2005cw,Mahmoudi:2009zx},
\begin{eqnarray}
 &&\frac{-1}{\sqrt{2}} \sum_{F=U,\,D,\,E}
 \bar{F}\Bigl\{  \left[ \kappa^Fs_{\beta-\alpha}+\rho^F c_{\beta-\alpha} \right] h^0 +
 \left[ \kappa^Fc_{\beta-\alpha}-\rho^Fs_{\beta-\alpha} \right] H^0 - i \, {\rm sgn}(Q_F)\rho^F A^0 \Bigr\} R F \nonumber \\
&&\quad\quad\quad\quad\quad\ \
 - \bar{U} \left[ V \rho^D R - \rho^{U\dagger} V L \right] D H^+
  -\bar{\nu} \left[ \rho^E R \right] E H^+ + {\rm H.c.} \,
\end{eqnarray}
where ${L,\,R} \equiv ( 1\mp \gamma_5 )/2$,
$c_{\beta-\alpha} = \cos(\beta-\alpha)$,
$s_{\beta-\alpha} = \sin(\beta-\alpha)$,
$\tan\beta \equiv v_2/v_1$, and
$\alpha$ is the mixing angle between neutral Higgs scalars, in the
notation~\cite{Gunion:1989we} of 2HDM-II.
The $\kappa$~matrices are diagonal and fixed by
fermion masses, $\kappa^F = \sqrt{2}m_F/v$ with $v \simeq 246$~GeV,
while $\rho$ matrices are in general not diagonal.
The off diagonal elements of $\rho$ are tree level FCNH couplings.
However, in the exact alignment limit of $c_{\beta - \alpha} = 0$,
the $h^0$ boson approaches SM Higgs and couples diagonally,
but $H^0$ and $A^0$ can still lead
to $\phi^0 \to \tau \mu$ at tree level.


As mentioned, there has been a lot of interest in
this FCNH interaction among leptons at the LHC.
There was a $2.4\sigma$ excess of
$h^0 \to \tau \mu$ above the background in CMS Run 1 data,
with the best fit branching fraction~\cite{Khachatryan:2015kon}
${\cal B}(h^0 \to \tau\mu) \simeq (0.84\pm 0.38)\%$, which is
consistent with ATLAS Run 1 result~\cite{Aad:2016blu} of
${\cal B}(h^0 \to \tau\mu) \simeq (0.77\pm 0.62)\%$.
But the excess was ruled out by 2016 CMS data~\cite{Sirunyan:2017xzt},
with upper limit $B(h^0 \to \tau\mu) \alt 0.25\%$,
giving the bound on FCNH coupling
$\sqrt{ |Y_{\tau\mu}|^2 + |Y_{\mu\tau}|^2 }
 = \tilde{\rho}_{\tau\mu}|c_{\beta -\alpha}| < 1.43 \times 10^{-3}$,
where
$\tilde{\rho}_{\tau\mu} \equiv
 \sqrt{ ( | \rho_{\tau\mu} |^2 + | \rho_{\mu\tau} |^2 ) / 2 }$.
However, $Y_{\tau\mu} = \rho_{\tau\mu}\,c_{\beta -\alpha}/\sqrt{2}$
may be small because of alignment, or $c_{\beta-\alpha} \to 0$.
The leptonic FCNH Yukawa couplings of the heavy $H^0$
boson, $Y'_{\tau\mu} = -\rho_{\tau\mu} \,s_{\beta-\alpha}/\sqrt{2}$
would approach the $A^0$ FCNH coupling in strength in the alignment limit,
since $s_{\beta-\alpha} \to 1$.
While the recent CMS limit implies $B(h^0 \to \tau\mu)$ must be small,
$B(H^0 \to \tau\mu)$ and $B(A^0 \to \tau\mu)$
can still be sizable and should be probed experimentally.

In this paper, we study the discovery potential for the decays
$H^0$, $A^0 \to \tau^{\pm} \mu^{\mp}$, followed by $\tau$ decays into
an electron and neutrinos or into a $\tau$-jet ($\pi,\rho$, or $a_1$)
and neutrino. Imposing the current LHC Higgs data, CMS and B physics
constraints, we calculate the full tree level matrix elements for both
signals and backgrounds. We use realistic acceptance cuts to
reduce the backgrounds with current b-tag, $\tau-$tag, and mistag
efficiencies. Some promising results are presented for 14 and 27 TeV
center of mass (CM) energies for an integrated luminosity
$\mathcal{L}$ = 300 and 3000 fb$^{-1}$,
in sync with future High Luminosity (HL) and High Energy (HE)
LHC~\cite{Barletta:2013ooa,Tomas:2016kuo,Zimmermann:2017bbr,Shiltsev:2017tjx}.

We discuss experimental limits on relevant parameters from B physics and
LHC Higgs data in Sec. II, and give in Sec. III
the production cross sections for the Higgs signal
and the dominant background with realistic acceptance cuts,
as well as our strategy to determine the reconstructed masses
for the Higgs bosons.
Sec. IV presents the discovery potential at the LHC
for $\sqrt{s} = 14$ TeV, and also for
future hadron colliders with $\sqrt{s} = 27$ and $100$ TeV.
Optimistic conclusions are drawn in Sec. V.

\section{Constraints on Relevant Parameters}

The most relevant parameters are $\rho_{\tau\mu}$, $\rho_{\mu\tau}$
for the decay $H^0/A^0 \to \tau \mu$, and $\rho_{tt}$ for the production
$gg \to H^0/A^0$ via the triangle-top loop.
A potentially large $\rho_{tc}$ induces~\cite{Altunkaynak:2015twa}
$H^0/A^0 \to t \bar c,~c \bar t$,
which can dilute the $H^0/A^0 \to \tau \mu$ branching ratio,
while $\rho_{ct}$ is subject to tight constraints by $B$ physics data.
LHC data for the 125 GeV $h$ boson~\cite{Sirunyan:2018koj,ATLAS:2019slw} 
suggest $|\cos(\beta - \alpha)| \ll 1$ in 2HDM-II.
We take $\cos(\beta - \alpha) = 0.1$ for illustration, although larger values
are allowed in the general 2HDM~\cite{Altunkaynak:2015twa,Hou:2018uvr}.
As for other $\rho$ matrix elements,
we set $\rho_{ff} = \kappa_f =\sqrt{2}m_f / v$
for diagonal elements except $\rho_{tt}$, and ignore off-diagonal ones
except $\rho_{\tau\mu}$, $\rho_{\mu\tau}$ and $\rho_{tc}$.
Degenerate extra scalar masses, i.e. $M_{H^0} = M_{A^0} = M_{H^\pm}$,
is assumed for simplicity.
In this section, we consider phenomenological constraints on $\rho_{\tau\mu}$,
$\rho_{\mu\tau}$, $\rho_{tt}$ and $\rho_{tc}$ under these assumptions.
In general $\rho_{tt}$ is complex and it may contribute to CP violation
and Baryogenesis~\cite{Fuyuto:2017ewj}.
For simplicity, we will take it to be real in this work.

%
%

In our analysis, we have set $\cos(\beta - \alpha) =$ 0.1 for case studies.
This choice of $\cos(\beta-\alpha)$ leads to cross sections of 
$pp \to H^0 \to W^+ W^- +X$ below current ATLAS limits~\cite{Aaboud:2017fgj} 
and it is consistent with recent LHC measurements for the light Higgs boson 
($h^0$)~\cite{Sirunyan:2018koj,ATLAS:2019slw}.
In Table I, we present cross sections of the heavier Higgs boson ($H^0$) 
decaying into a pair of $W$ bosons in general two Higgs doublet models 
at $\sqrt{s} = 14$ TeV 
and experimental limits from ATLAS with $\sqrt{s}$ = 13 TeV.

\begin{table}[htb]
\begin{center}
\begin{tabular}{|p{2cm}|p{2.8cm}|p{2.8cm}|p{4cm}|}
\hline
$M_H$ (GeV)& $\lambda_5$ = 0 \; (fb) &   $\lambda_5$ = -1 \; (fb) &
ATLAS limit \; (fb)~\cite{Aaboud:2017fgj} \\ 
\hline
 300   & 1.23$\times 10^3$ & 1.98$\times 10^3$ & $\leq$ 8.00$\times 10^3$ \\
\hline
 400   & 7.17$\times 10^2$ & 9.49$\times 10^2$ & $\leq$ 1.30$\times 10^3$ \\ 
\hline
 500   & 2.17$\times 10^2$ & 2.47$\times 10^2$ & $\leq$ 4.00$\times 10^2$ \\ 
\hline
\end{tabular}
\caption{Cross section of $pp \to W^+ W^- +X$ at $\sqrt{s} = 14$ TeV 
and ATLAS limits at $\sqrt{s} = 13$ TeV.}
\label{ppWW}
\end{center}

\end{table}

%
%

The FCNH couplings $\rho_{\tau\mu}$ and $\rho_{\mu\tau}$ induce
$h^0 \to \tau \mu$ decay, with branching ratio
\begin{align}
\mathcal{B}(h^0 \to  \tau\mu) = \frac{M_{h^0} c_{\beta -\alpha}^2}{16\pi \Gamma_{h^0}}
(|\rho_{\tau \mu}|^2 + |\rho_{\mu \tau}|^2),
\end{align}
where $M_{h^0} \simeq 125$~GeV, and the $\tau^+ \mu^-$ and $\mu^+ \tau^-$
modes are added up.
The total width $\Gamma_{h^0}$ is estimated by the sum of
$h^0 \to WW^*, ZZ^*, gg, b\bar b, c\bar c$ and $\tau^+\tau^-$ partial widths
obtained by rescaling of SM values~\cite{deFlorian:2016spz}
with $\Gamma(h^0 \to \tau\mu)$ added.
We impose the 95\% C.L. limit $\mathcal{B}(h^0 \to \tau \mu) < 0.25$\%
by CMS~\cite{Sirunyan:2017xzt}.

Constraints on $\rho_{\tau\mu}$ and $\rho_{\mu\tau}$ by various low-energy
processes containing tau and muon are discussed in the literature (see, e.g.
Ref.~\cite{Davidson:2010xv,Sierra:2014nqa,Dorsner:2015mja,Omura:2015xcg}).
It is found that $\tau \to \mu \gamma$ is most relevant.
Its branching ratio is given by~\cite{Omura:2015xcg}
\begin{align}
{\cal B}(\tau\to \mu\gamma)
= \frac{48\pi^3\alpha}{G_F^2}(|A_L|^2+|A_R|^2){\cal B}(\tau\to\mu\bar\nu_\mu\nu_\tau),
\end{align}
where we take
${\cal B}(\tau\to\mu\bar\nu_\mu\nu_\tau)=(17.39 \pm 0.04)$\%~\cite{Tanabashi:2018oca},
and $A_{L,R}$ gives the strength of the $\tau \to \mu\gamma$ amplitude
with different chiral structure.
In addition to the one-loop contribution mediated by the neutral and charged
scalar bosons, we also include the two-loop Barr-Zee type contribution in $A_{L,R}$,
following Ref.~\cite{Omura:2015xcg}.
The latter contribution can be obtained by the obvious translation of the expression
for $\mu \to e \gamma$~\cite{Chang:1993kw},
and we include the dominant contribution from the effective
$\phi^0 \gamma\gamma$ ($\phi^0 = h^0, H^0, A^0$) vertex,
which brings in dependence on $\rho_{tt}$ via the top loop.
Current limits on $\tau \to \mu \gamma$
are $\mathcal{B}(\tau \to \mu \gamma) < 4.5 \times 10^{-8}$
by Belle~\cite{Hayasaka:2007vc} and $4.4 \times 10^{-8}$
by BABAR~\cite{Aubert:2009ag}, both at 90\% C.L.
Belle II may improve the limit by a factor of 100~\cite{Kou:2018nap}.
We conservatively take $\mathcal{B}(\tau \to \mu \gamma) = 10^{-9}$
to illustrate future sensitivity.

$\rho_{tt}$ is also constrained by $B$ physics, in particular by the
$B_q$ ($q=d,s$) meson mixings and $b \to s\gamma$~\cite{Altunkaynak:2015twa}.
We update the results of Ref.~\cite{Altunkaynak:2015twa} with the latest
experimental and theoretical values as summarized below.
We adopt the Summer 2018 result by UTfit \cite{Bona:2006sa}
for values of CKM parameters and constraints on the $B_q$-$\bar B_q$
mixing amplitude ($M_{12}^q$):
\begin{align}
C_{B_d}&\in [0.83,~1.29], \quad \phi_{B_d}\in [-6.0^\circ,~1.5^\circ], \notag\\
C_{B_s}&\in [0.942,~1.288], \quad \phi_{B_s}\in [-1.35^\circ,~2.21^\circ] \quad
{\rm at~95\%~probability,}
\label{eq:UTfit}
\end{align}
where $C_{B_q}e^{2i\phi_{B_q}}\equiv M_{12}^q/M_{12}^q|_{\rm SM}$.
As for $b\to s\gamma$, we adopt a recent world average
$B(\bar B\to X_s \gamma)_{\rm exp} = (3.32\pm 0.15) \times 10^{-4}$~\cite{Amhis:2016xyh},
which includes the recent Belle result~\cite{Belle:2016ufb},
and the updated SM prediction $B(\bar B\to X_s \gamma)_{\rm SM} = (3.36\pm 0.23)
\times 10^{-4}$~\cite{Misiak:2015xwa,Czakon:2015exa}
for the photon energy $E_\gamma > 1.6$ GeV.
We then use the ratio~\cite{Crivellin:2013wna}
$R^{b\to s\gamma}_{\rm exp} = \mathcal{B}(\bar B\to X_s \gamma)_{\rm exp} /
\mathcal{B}( \bar B\to X_s \gamma)_{\rm SM}$ to constrain
$R^{b\to s\gamma}_{\rm theory}= \mathcal{B}( \bar B\to X_s \gamma)_{\rm 2HDM}/
\mathcal{B}(\bar B \to X_s \gamma)_{\rm SM}$ based on our LO calculation,
allowing the 2$\sigma$ experimental uncertainty of $R^{b\to s\gamma}_{\rm exp}$
with the theoretical uncertainty linearly added.
Note that the new experimental and theoretical values result in
rather strong limits on $M_{H^\pm}$ in 2HDM-II~\cite{Misiak:2017bgg}:
$M_{H^\pm} >$ 570--800 GeV at 95\% CL, depending on the method
used to extract the limit. Our method gives $M_{H^\pm} \gtrsim 710$ GeV
in 2HDM-II at large $\tan\beta$ 

We ignore effect of $\rho_{tc}$ on $B_q$ mixings and $b \to s\gamma$
as it enters via the charm loop, making its impact minor~\cite{Crivellin:2013wna}
compared with $\rho_{tt}$ and $\rho_{ct}$ entering via the top loop.
But $\rho_{tc}$ induces $t \to ch$ decay~\cite{Hou:1991un,Chen:2013qta}, and
the recent ATLAS limit~\cite{Aaboud:2018oqm} of
\begin{eqnarray}
\mathcal{B}(t \to ch^0) < 1.1 \times 10^{-3}\ (95\%\ {\rm C.L.})
\end{eqnarray}
directly constrains $\rho_{tc}$ if $c_{\beta -\alpha}$ is nonzero.
The $t \to ch^0$ width is given by
\begin{eqnarray}
\Gamma(t\to ch^0)
& = & \frac{m_t \, c_{\beta -\alpha}^2}{32\pi}\lambda^{1/2}(1,x_c,x_h)
 \left[ (1+x_c-x_h)\frac{|\rho_{tc}|^2+|\rho_{ct}|^2}{2}
 +2\sqrt{x_c}\, {\rm Re}(\rho_{tc}\rho_{ct}) \right] \nonumber \\
& \simeq & \frac{m_t \, c_{\beta -\alpha}^2\,\tilde{\rho}_{tc}^2}{32\pi}
 \times \left( 1-x_h \right)^2 \, ,
\end{eqnarray}
where $\lambda(x,y,z) = x^2+y^2+z^2-2xy-2yz-2zx$, $x_c= m_c^2/m_t^2$,
$x_h= M_h^2/m_t^2$, and we define
$\tilde{\rho}_{tc} = \sqrt{|\rho_{tc}|^2+|\rho_{ct}|^2}/\sqrt{2}$
as a convenient FCNH coupling~\cite{Altunkaynak:2015twa,Jain:2019ebq}.
Combining with the LO $t\to bW$ width to obtain the total top width,
we recast the ATLAS limit~\cite{Aaboud:2018oqm} of Eq.~(5) to obtain
$\lambda_{tch} = |\tilde{\rho}_{tc} \, c_{\beta -\alpha}|
 = |\rho_{tc} \, c_{\beta -\alpha}|/\sqrt{2} \lesssim 0.064$
for $\rho_{ct} = 0$.

\begin{figure}[htb]
\centering
{\includegraphics[width=70mm]{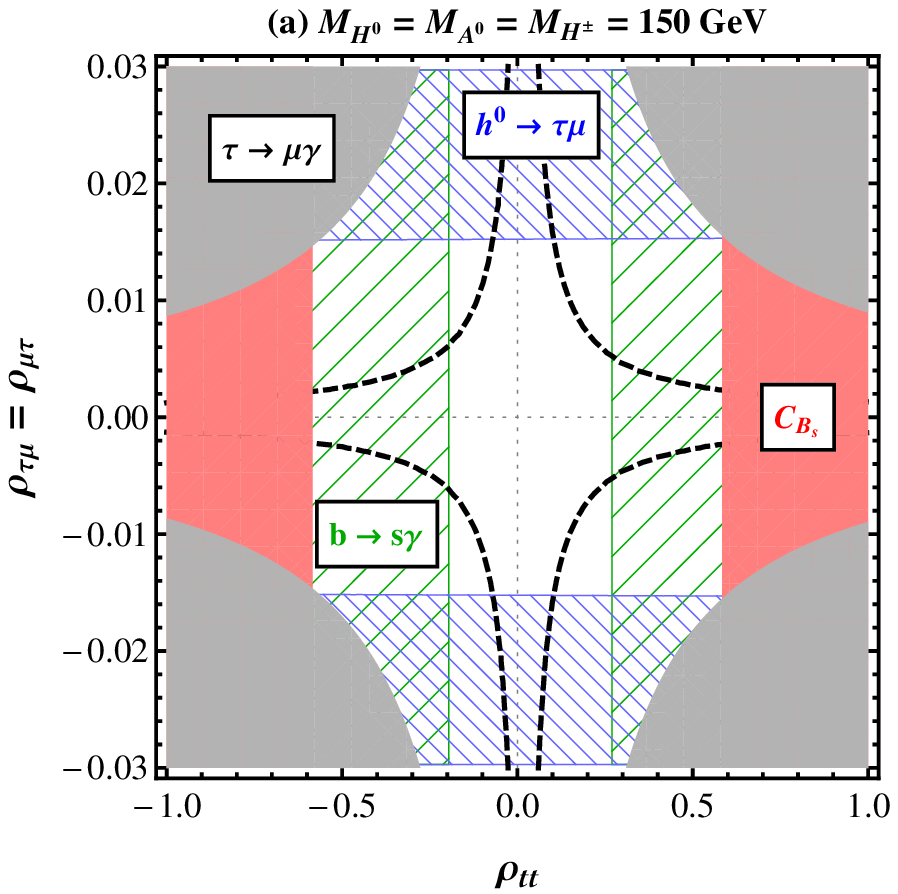}
 \hspace{0.1in}
 \includegraphics[width=70mm]{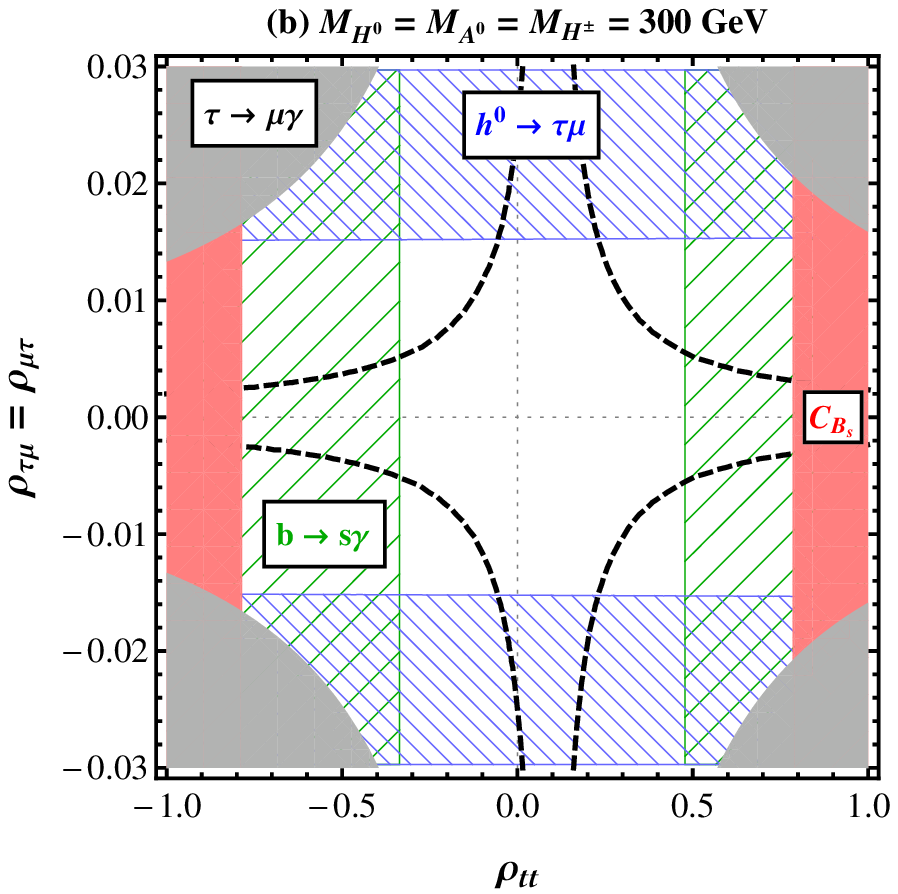}
}
\caption{
Constraints on the ($\rho_{tt},\rho_{\tau\mu}$) plane with
$\rho_{\mu\tau} = \rho_{\tau \mu}$ for
(a) $M_{H^0} = M_{A^0} = M_{H^\pm} = 150$ GeV and (b) 300 GeV,
both assuming $\cos(\beta -\alpha) = 0.1$, $\rho_{\tau\tau}=\kappa_\tau$
and $\rho_{bb} = \kappa_b$.
Dashed lines indicate $\mathcal{B}(\tau \to \mu\gamma) = 10^{-9}$
for a future sensitivity. See the main text for details.
} \label{fig:flavor}
\end{figure}

In our numerical calculations of this section,
we take the latest PDG values~\cite{Tanabashi:2018oca}
for particle masses, in particular the top quark pole mass $m_t = (173.1 \pm 0.9)$ GeV
and bottom quark $\overline{\rm MS}$ mass $m_b(m_b) = (4.18^{+0.04}_{-0.03})$ GeV
as input.
Fig.~\ref{fig:flavor} summarizes the constraints on
the $(\rho_{tt}, \tilde\rho_{\tau\mu})$ plane with $\rho_{\mu\tau}=\rho_{\tau\mu}$
for (a) $M_{H^0} = M_{A^0} = M_{H^\pm} =$ 150 GeV and (b) 300 GeV:
exclusions are shown by
the blue-hatched regions for $h^0 \to \tau \mu$ by CMS,
gray-shaded regions for $\tau \to \mu\gamma$ by BABAR,
pink-shaded regions for the $B_s$ mixing ($C_{B_s}$) and
green-hatched regions for $b \to s\gamma$.
The other three observables in Eqs.~(\ref{eq:UTfit}) give weaker limit than $C_{B_s}$
and are not shown in the figures.
The dashed contours with $\mathcal{B}(\tau \to \mu \gamma) = 10^{-9}$
are shown as future Belle II sensitivity.
We note that the constraints by $h^0 \to \tau \mu$ and $b \to s\gamma$
are highly sensitive to the choice of parameters:
the $h^0 \to \tau \mu$ constraint gets weaker for a smaller $c_{\beta -\alpha}$
and eventually loses sensitivity if $c_{\beta -\alpha} = 0$;
the $b \to s\gamma$ constraint is relaxed for a smaller $|\rho_{bb}|$,
and becomes weaker than the $B_s$ mixing constraint if $\rho_{bb}=0$.
In passing, the effect~\cite{Omura:2015xcg} on the muon $g-2$ is insignificant
($|\delta a_\mu| < \mathcal{O}(10^{-12})$ in the shown parameter regions)
due to small $\rho_{\tau\mu}/\rho_{\mu\tau}$ and $c_{\beta -\alpha}$ values,
which suppress the one-loop $h^0$ contribution, and the $H^0$--$A^0$ mass degeneracy,
which leads to cancellation of the one-loop $H^0$ and $A^0$ contributions.

Combining experimental limits from LHC Higgs data and $B$ physics,
we consider
$\rho_{\tau\mu} = \rho_{\mu\tau} < 0.01$, and
$|\tilde\rho_{tc}\,c_{\beta -\alpha}|
 = \lambda_{tch} < 0.064$~\cite{Aaboud:2018oqm}.
To be consistent with $B$ physics constraints, we choose
\begin{eqnarray}
\rho_{tt} = 0.2 \times (M_\phi / 150 \, {\rm GeV}),
\label{eq:rho_tt}
\end{eqnarray}
for $\phi^0 = H^0$ or $A^0$,
which always satisfies the $b \to s \gamma$
  constraint for the heavy Higgs scalar mass considered in our study.

\section{Higgs Signal and Physics Background}

In this section, we discuss the prospect of discovering FCNH interactions
from heavy Higgs bosons $H^0$ and $A^0$ decaying into $\tau^\pm\mu^\mp$.
There are several parameters that can affect the signal cross section
in the 2HDM. We use the experimental results and constraints to
optimize the parameter range.
Recent data from LHC point toward a Higgs sector in which the light
CP even Higgs state is the SM-like Higgs~\cite{Sirunyan:2018koj,ATLAS:2019slw}.
This constraint suggests that $c_{\beta -\alpha}$ is very small.
For case studies in our analysis we set $c_{\beta -\alpha} = 0.1$.

\subsection{The Higgs Potential and Decay Final States}

For the heavy CP-even $H^0$ boson, the most important SM
decay channels are $b\bar{b}, t\bar{t}, WW$, and $ZZ$.
In addition, $tc$ and $h^0h^0$ channels might become dominant
in some regions of parameter space.
The CP-odd pseudoscalar  $A^0$ boson has significant decays into
$b\bar{b}, t\bar{t}$, as well as
possible dominant contributions from $tc$ and $Z h^0$ channels.

\begin{figure}[b]
\begin{center}
\includegraphics[width=72mm]{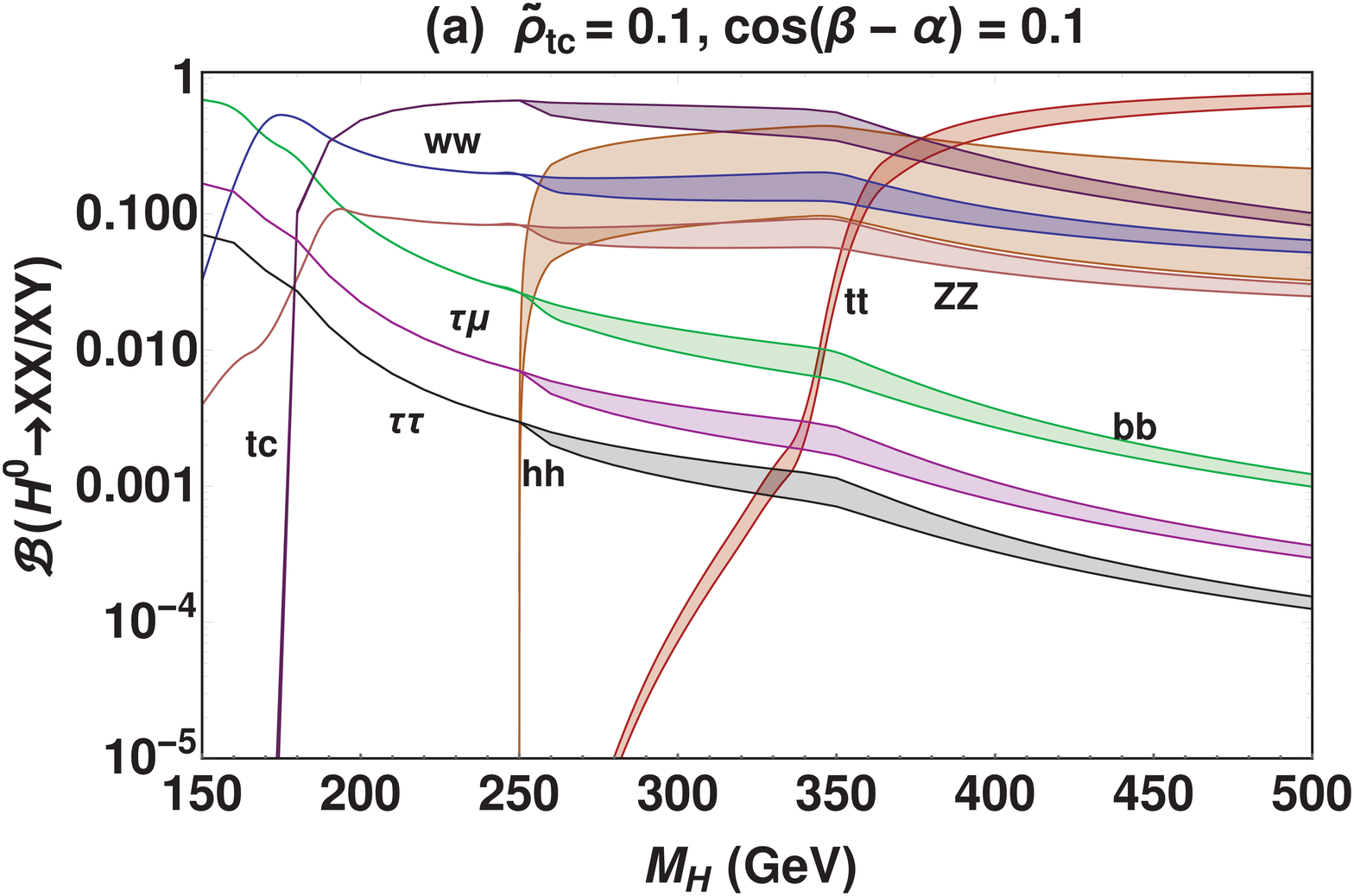}
\hspace{0.1in}
\includegraphics[width=72mm]{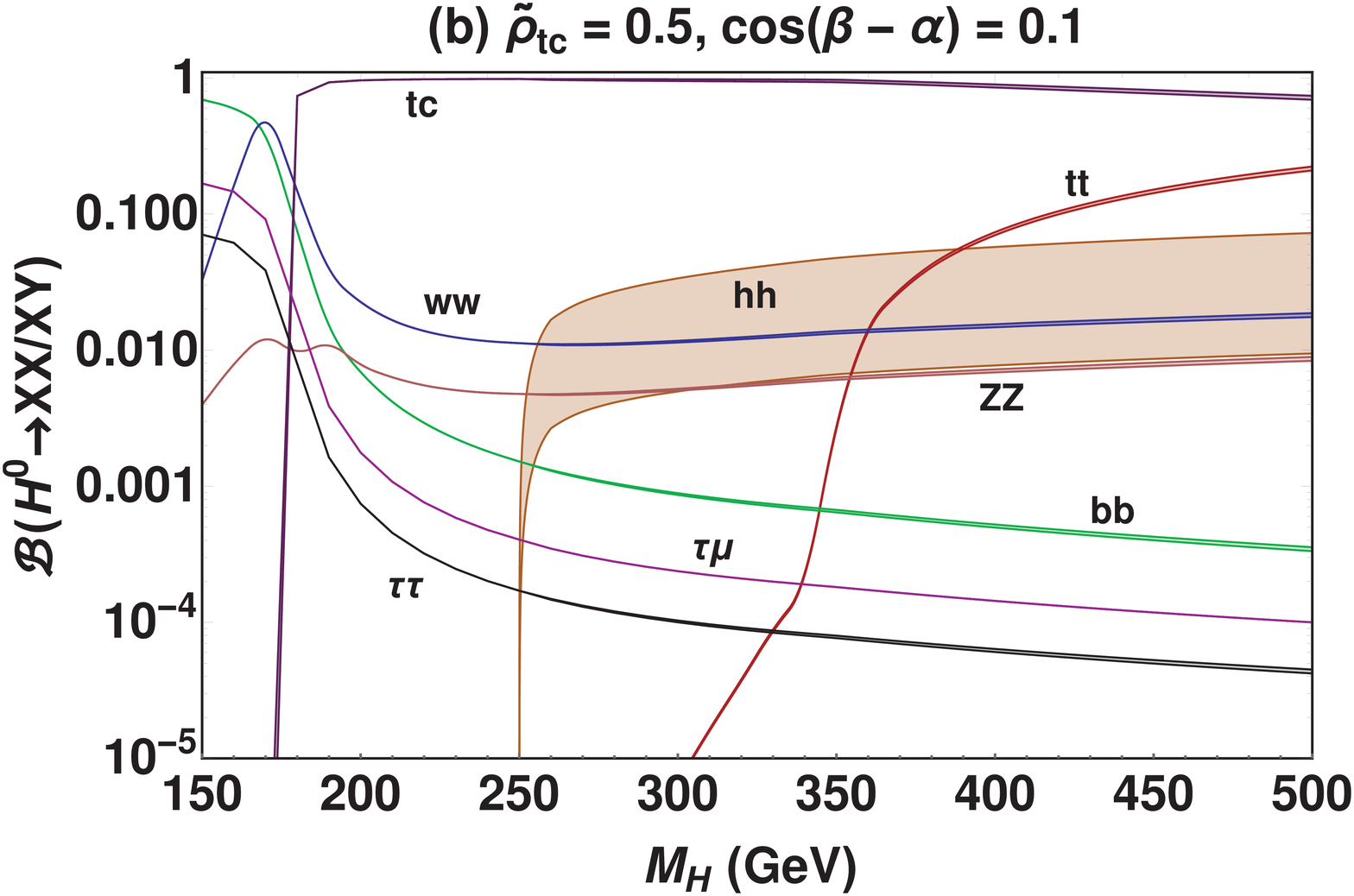} \\
\includegraphics[width=72mm]{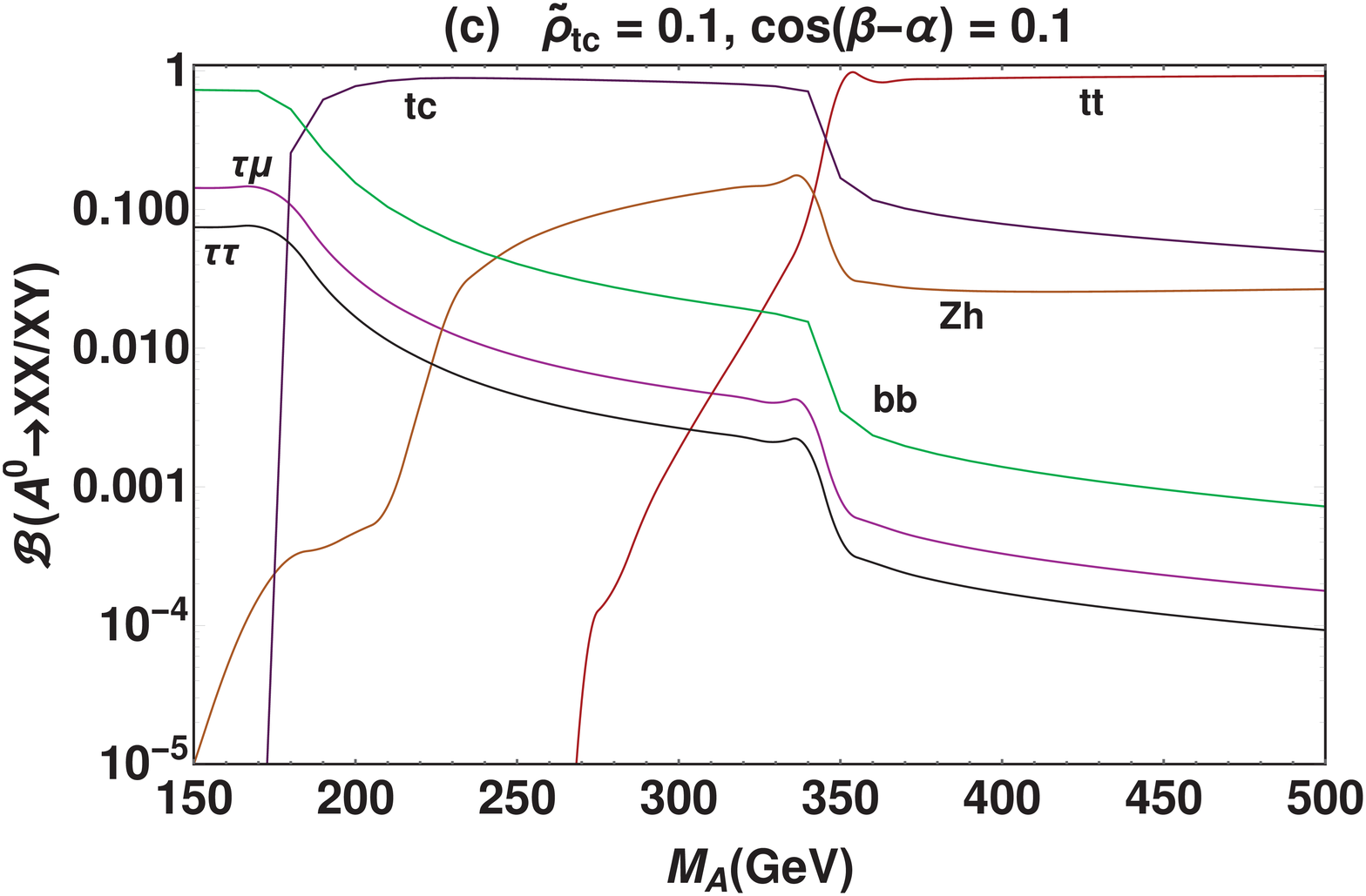}
\hspace{0.1in}
\includegraphics[width=72mm]{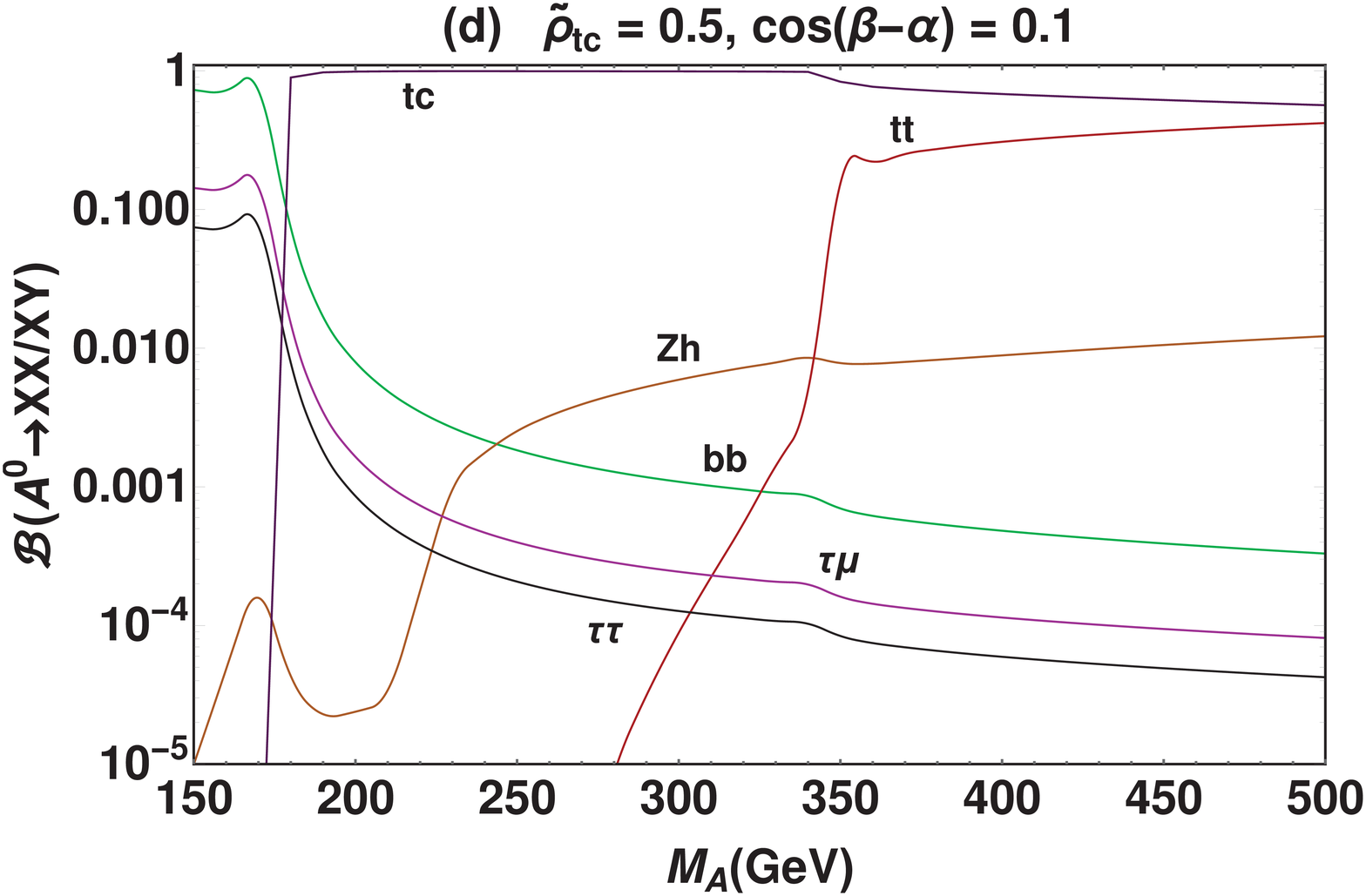}
\caption{Major two body decays of $H^0$ vs $M_H$
for (a) $\tilde{\rho}_{tc} = 0.1$, and (b) $\tilde{\rho}_{tc} = 0.5$,
with $\tilde{\rho}_{\tau \mu} = 0.01$, $\lambda_5 = 0$ and
1 $\leq$ tan$\beta$ $\leq$ 10.
The analogous case for $A^0$ is given in (c) and (d).
} \label{fig:BFHA}
\end{center}
\end{figure}

To study heavy boson $H^0$ or $A^0$ decays involving the
light Higgs boson $h^0$, let us consider a general CP-conserving
Higgs potential~\cite{Gunion:2002zf}
\begin{eqnarray}
\nonumber
V &=& m_{11}^2 |\Phi_1|^2 + m_{22}^2 |\Phi_2|^2 - \left[m_{12}^2 \Phi_1^\dag \Phi_2 + {\rm h.c.} \right] +\frac{1}{2} \lambda_1 |\Phi_1|^4 + \frac{1}{2} \lambda_2 |\Phi_2)|^4 + \lambda_3 |\Phi_1|^2|\Phi_2|^2 \\ \label{eq:potential}
&+& \lambda_4 (\Phi_1^\dag \Phi_2) (\Phi_2^\dag \Phi_1)
+ \left[ \frac{1}{2} \lambda_5 (\Phi_1^\dag \Phi_2)^2 + \lambda_6 |\Phi_1|^2 (\Phi_1^\dag \Phi_2) + \lambda_7 |\Phi_2|^2 (\Phi_1^\dag \Phi_2) + {\rm h.c.} \right]
\label{Higgs_potential}
\end{eqnarray}
Applying minimization conditions, we can express the triple Higgs
coupling $g_{Hhh}$ in terms of physical masses and
mixing angles~\cite{Gunion:2002zf,Craig:2013hca}
\begin{eqnarray}
g_{Hhh} \simeq - \frac{c_{\beta - \alpha}}{v}
 & \bigg[ &  4 m_A^2 - 2m_h^2 - m_H^2  + 4 \lambda_5 v^2 \nonumber \\
& & + \frac{2 v^2}{\tan2 \beta} (\lambda_{6} - \lambda_7)
    + \frac{2 v^2}{\sin2 \beta} (\lambda_{6} +\lambda_{7}  )
    + {\cal O}(c_{\beta - \alpha}) ~ \bigg] \, .
\end{eqnarray}
which vanishes in the alignment limit, as the self coupling is proportional to
$c_{\beta - \alpha}$.

For simplicity, we take the heavy Higgs states $H^0,A^0$ and $H^{\pm}$
to be degenerate 
and we set $\lambda_{6,7} = 0$.
As a sample study, we choose three values of $\lambda_5 = \pm 1$ and 0,
to maintain tree level unitarity.
For Yukawa couplings, except $\rho_{tt}$ (Eq.~\ref{eq:rho_tt}),
we set $\rho_{ii} = \kappa_{i}$, which is in good agreement
with the current constraints from B Physics and LHC.
For off-diagonal elements $\rho_{ij}$, we perform case studies
for $\tilde{\rho}_{tc}$= 0.1 and 0.5, and set all the remaining
off-diagonal terms to be 0 except $\rho_{\tau \mu}$.

Figure 2 shows all major two body decays for
the heavy $H^0$ and $A^0$,
with $\tilde{\rho}_{tc} = 0.1$ and $0.5$.
Note that, in keeping $\rho_{\tau\mu} = \rho_{\mu\tau}$
as in Fig.~\ref{fig:flavor} and fixing the value to 0.01,
$H^0$, $A^0 \to \tau\mu$ dominates over $\tau\tau$,
which is interesting by itself.
For the $H^0$ boson,
$\tilde{\rho}_{tc}$, $\lambda_5$ and $\tan\beta$ play crucial roles in
affecting the $H^0 \to \tau \mu$ branching ratio.
We use 2HDMC~\cite{Eriksson:2009ws} to scan over
150 GeV $\leq M_H \leq$ 500 GeV
and $ 1 \leq \tan\beta \leq 10$ for $\lambda_5 = 0$.
For $M_H > 2 m_t$, $H^0 \to t \bar{t}$, $h^0 h^0$ and $t c$
channels might become predominant.
This suggests $M_H$ values close to 1 TeV may not be visible
in the $\tau\mu$ channels, so we limit our case study to $M_{H} < 500$ GeV.

The pseudoscalar $A^0$ decays mostly into fermions,
as shown in Figs. 2(c) and 2(d).
Its decay is independent of $\tan\beta$ and $\lambda_5$ in a general 2HDM.
Only $\rho_{tc}$ has significant impact on the branching fractions.
For $\tilde{\rho}_{tc} \gtrsim 0.5$, $A^0 \to t\bar{c}+\bar{t}c$ becomes dominant.
Furthermore, for $M_A > 220$ GeV, $A^0 \to Zh^0$ also makes significant
contribution.
For $M_A > 2 m_t$, the $t\bar t$ channel starts to dominate,
hence we limit our study to $M_A < 500$ GeV to ensure significance.

\subsection{Higgs Signal}

Our main signal channel is the production and FCNH decay of a
heavy Higgs boson ($\phi^0 = H^0, A^0$) via gluon fusion,
$pp \to \phi^0 \to \tau \mu +X$~\cite{
Han:2000jz,Assamagan:2002kf,Harnik:2012pb,
Buschmann:2016uzg,Sher:2016rhh,Primulando:2016eod,Bednyakov:2018hfq}.
With the $\tau$ decaying leptonically, we
are looking for a final state of two opposite sign, different flavor
leptons and missing energy.  With a hadronically decaying $\tau$, a
final state with a $\tau$-jet ($j_\tau$), a muon, and missing
energy is needed.
We have evaluated the FCNH signal cross sections with analytic matrix
element and leading order CT14 parton distribution
functions~\cite{Dulat:2015mca,Gao:2013xoa}.
To include higher order corrections we calculate K-factors with
Higlu~\cite{Spira} for $p p \to \phi^0 + X$.

\subsection{Standard Model Backgrounds}

The dominant background for leptonic final states comes from
$p  p \to \tau \tau \to e \mu + \notE_T + X$, $p p \to W^+ W^- +X$
and $p p \to h \to \tau \tau +X$.
For hadronic channel, we have considered
$p p \to W^{\pm} j \to \mu j +\notE_T + X$
as the most dominant background along with the
$\tau\tau$ channel.
For hadronic channel, $t\bar{t}$ contribution is highly suppressed,
when we veto any event with more than one b jet, with
$p_T > 20$ GeV and $|\eta| < 4.7$.
We have used MADGRAPH~\cite{Alwall:2011uj} and HELAS~\cite{Hagiwara:2008jb}
to generate matrix elements for the backgrounds.
To include the NLO corrections,
we have employed MCFM~\cite{Campbell:2010ff,Campbell:2015qma}
to evaluate higher order cross sections.

\begin{table}[t]
\begin{center}
 \begin{tabular}{|p{3.5cm}|p{3.5cm}|p{3.5cm}|}
 \hline
 {\bf Kinematic Variables} & $\phi^0 \to \tau\mu \to e \mu +X$
                           & $\phi^0 \to \tau\mu \to j_{\tau}\mu +X$ \\
 \hline
 $P_T(e)$ &  $ > 10\, {\rm GeV}$  &    \\
 $P_T(\mu)$ & $ > 26\, {\rm GeV}$  & $ >26\, {\rm GeV}$ \\
 $P_T(j_{\tau})$   &        &  $> 30\, {\rm GeV}$ \\
 \hline
 $|\eta_e|$ & $< 2.3 $ & \\
 $|\eta_{\mu}|$ & $ < 2.4$ & $ < 2.4$ \\
 $|\eta_{j_{\tau}}|$ & & $<2.3$ \\
 \hline
 $\Delta R(e,\mu)$ & $ > 0.3$ &  \\
 $\Delta R(j_{\tau},\mu)$ &  & $> 0.5$ \\
 $\Delta \phi(e,\vec{p}^{\rm miss}_{T})[{\rm radians}]$& $< 0.7$ & \\
 $\Delta \phi(e,\mu)[{\rm radians}]$ & $> 2.5$ & \\
 \hline
 $M_T(\mu,\notE_T)$ & $> 60\, {\rm GeV}$ &   \\
 $M_T(e,\notE_T)$ & $< 50\, {\rm GeV}$ & \\
 $M_T(j_{\tau},\notE_T)$  &  & $< 105\, {\rm GeV}$ \\
 $|M_{\rm col}(\tau\mu) - M_{\phi}|$ & $< 0.2\times M_{\phi}$
                        & $< 0.2\times M_{\phi}$ \\
 \hline
\end{tabular}
\caption{Acceptance cuts for
(a) the leptonic channel $\phi^0 \to \tau\mu \to e\mu +X$, and
(b) the hadronic channel $\phi^0 \to \tau\mu \to j_\tau \mu +X$, where
$\phi^0 = H^0$ or $A^0$.
} \label{tab:cuts}
\end{center}
\end{table}

\subsection{Realistic acceptance cuts}

To study the discovery potential for the FCNH signal, we apply realistic
acceptance cuts proposed by CMS~\cite{Khachatryan:2015kon,Sirunyan:2017xzt} at
$\sqrt{s} = 13$ TeV as shown in Table II.
In addition, we apply Gaussian smearing for particle
momenta~\cite{ATLAS:2013-004,Khachatryan:2016kdb}
to simulate detector effects based on ATLAS~\cite{Aad:2009wy} and
CMS~\cite{Colaleo:2015vsq} specifications.
\begin{equation}
\frac{\Delta E}{E} = \frac{0.60}{\sqrt{E\, {\rm (GeV)}}} \oplus 0.03\; ({\rm jets}) \, ,
 \quad
\frac{\Delta E}{E} = \frac{0.25}{\sqrt{E\, {\rm (GeV)}}} \oplus 0.01\; ({\rm leptons}) \, .
\end{equation}

We present in Table III the cross sections for physics backgrounds
with acceptance cuts as well as tagging efficiency for $\tau$-jets,
$\epsilon_{j_{\tau}}$ = 0.7~\cite{Friis:2011zz,Lumb:2010btk},
and mistag efficiency $\epsilon_{j}$ = 0.01~\cite{ATLAS:2018bpl,Sirunyan:2017ezt}.

\begin{table}[t]
 \begin{center}
 \begin{tabular}{|p{4.0cm}|p{1.5cm}|p{1.5cm}|p{1.5cm}|}
 \hline
  CM Energy ($\sqrt{s}$)  & 14 TeV & 27 TeV & 100 TeV \\
  \hline
  \multicolumn{4}{c}{\textbf{Backgrounds for $\tau\mu \to e\mu + X$}} \\
  \hline
  $p p \to \tau \tau + X$ & 31.96 & 58.7 & 195.1 \\
  \hline
  $p p \to W^+ W^- + X$   & 12.27 & 23.73 & 86.29 \\
  \hline
  $p p \to h^0 \to \tau \tau + X$ & 1.92 & 5.06 & 27.9 \\
  \hline
  $p p \to h^0 \to W^+ W^- + X$   & 0.95 & 2.51 & 13.9 \\
  \hline
  \textbf{Total}                  & 47.1 & 90.0 & 323.2 \\
  \hline
  \multicolumn{4}{c}{\textbf{Backgrounds for $\tau\mu \to j_{\tau}\mu + X$}} \\
  \hline
  $p p \to W^{\pm} j +X$  & 6139 & 14074 & 61285 \\
  \hline
  $p p \to \tau \tau +X$  & 109.8 & 202.3 & 676.9 \\
  \hline
  $p p \to h^0 \to \tau\tau +X$ & 6.4 & 16.9 & 93.3 \\
  \hline
  \textbf{Total}                & 6255 & 14293 & 62055 \\
  \hline
 \end{tabular}
\caption{
Physics background cross sections in fb for
(a) leptonic $\phi^0 \to \tau\mu \to e\mu +X$, and
(b) hadronic $\phi^0 \to \tau\mu \to j_\tau \mu +X$,
with $M_{\phi} = 125.1$ GeV, for
$\sqrt{s} = 14$, 27, and 100 TeV.
We apply $\tau$-jet tagging efficiency
$\epsilon_{j_{\tau}}$ = 0.7~\cite{Friis:2011zz,Lumb:2010btk},
and mistag efficiency for other jets
$\epsilon_{j}$ = 0.01~\cite{ATLAS:2018bpl,Sirunyan:2017ezt}.
} \label{tab:sigma_b}
 \end{center}
\end{table}

We note that, as the Higgs boson mass increases,
$M_{\rm col}(\tau\mu)$ cut becomes more effective,
and for $M_H > 180\, {\rm GeV}$,
$p p \to h^0 \to \tau \tau +X$,
$p p \to h^0 \to W^+ W^- +X$ are almost completely vetoed.
For leptonic channel,
$p p \to W^+ W^- +X$ becomes more dominant than
$p p \to \tau \tau +X$.

\section{Discovery Potential}

To estimate the discovery potential, we require that
the lower limit on the signal plus background should be larger than
the corresponding upper limit on the background
with statistical fluctuations, which leads to~\cite{HGG}
\begin{equation}
    \sigma_S \ge \frac{N}{\cal L} \left[ N + 2\sqrt{{\cal L}\,\sigma_{B}} \right]
\end{equation}
where $\sigma_S$ and $\sigma_B$ are the signal and background
cross sections, respectively,  and ${\cal L}$ is the integrated luminosity.
Choosing $N = 2.5$, we obtain a $5\sigma$ significance.
For a large number of background events,
it simplifies to the statistical significance
 \begin{equation}
  N_{SS} = \frac{N_S}{\sqrt{N_B}} = \frac{{\cal L}\,\sigma_S}{\sqrt{{\cal L}\,\sigma_B}}
  \geq 5 \, ,
 \label{eq:NSS}
 \end{equation}
where $N_S$ and $N_B$ are the number of signal and background events.

%
%

To show sensitivity of possible systematic uncertainties on the Higgs signal of
$H^0 \to \tau\mu$, we present the ratio of $N_S/N_B$ as well as
statistical significance $N_{SS} = N_S\sqrt{N_B}$ in Table IV, where 
$N_S =$ number of signal events and $N_B =$ background events for 
(a) $\tau \to e$ and (b)$\tau \to j_{\tau}$ 
with $\rho_{\tau \mu} =$ 0.005 and $\lambda_5=$ 0 
at $\sqrt{s} = 14$ TeV and $\mathcal{L}$ = 3000 $fb^{-1}$. 
These data give the totality of the relevant information 
pertaining to the strength of the signal versus background. 
Note that the cross section of CP-odd pseudoscalar ($A^0$) will be
larger than that of ($H^0$).  

\begin{table}[htb]
\begin{center}

\begin{tabular}{|p{2cm}|p{2cm}|p{2cm}|p{2cm}|p{2cm}|p{3cm}|}
 \multicolumn{5}{c}{\bf (a) $pp \to H^0 \to \tau\mu \to e\mu + X$} \\
\hline 
$M_H$(GeV) & Total Background & Signal & Min(for 5 $\sigma$) &
$N_S/N_B$ & $N_{SS} \equiv N_S/\sqrt{N_B}$   \\ \hline 
150 & 1.06$\times 10^5$ & 4.68$\times 10^3$ & 1.63$\times 10^3$ & 0.044 & 14.4 \\ 
\hline
200 & 7.73$\times 10^4$ & 1.29$\times 10^3$ & 1.39$\times 10^3$ & 0.017 & 4.66 \\
\hline
300 & 4.47$\times 10^4$ & 3.94$\times 10^2$ & 1.06$\times 10^3$ & 0.009 & 1.86 \\
\hline
400 & 2.51$\times 10^4$ & 2.74$\times 10^2$ & 7.99$\times 10^2$ & 0.011 & 1.73 \\
\hline
\multicolumn{5}{c}{\bf (b) $pp \to H^0 \to \tau\mu \to j_\tau \mu + X$} \\
\hline 
150 & 1.51$\times 10^7$ & 1.54$\times 10^4$ & 1.94$\times 10^4$ & 0.001  & 3.95 \\ \hline 
200 & 7.50$\times 10^6$ & 4.20$\times 10^3$ & 1.37$\times 10^4$ & 0.0006 & 1.55\\ \hline  
300 & 1.90$\times 10^6$ & 1.29$\times 10^3$ & 6.89$\times 10^3$ & 0.0007 & 0.94\\ \hline  
400 & 5.85$\times 10^5$ & 8.88$\times 10^2$ & 3.83$\times 10^3$ & 0.0015 & 1.16\\ 
\hline
 
\end{tabular}

\caption{Comparison of number of signal events versus number of
  background events.}
\label{ns-nb}

\end{center}
\end{table}

\begin{figure}[b!]
\begin{center}

\includegraphics[width=78mm]{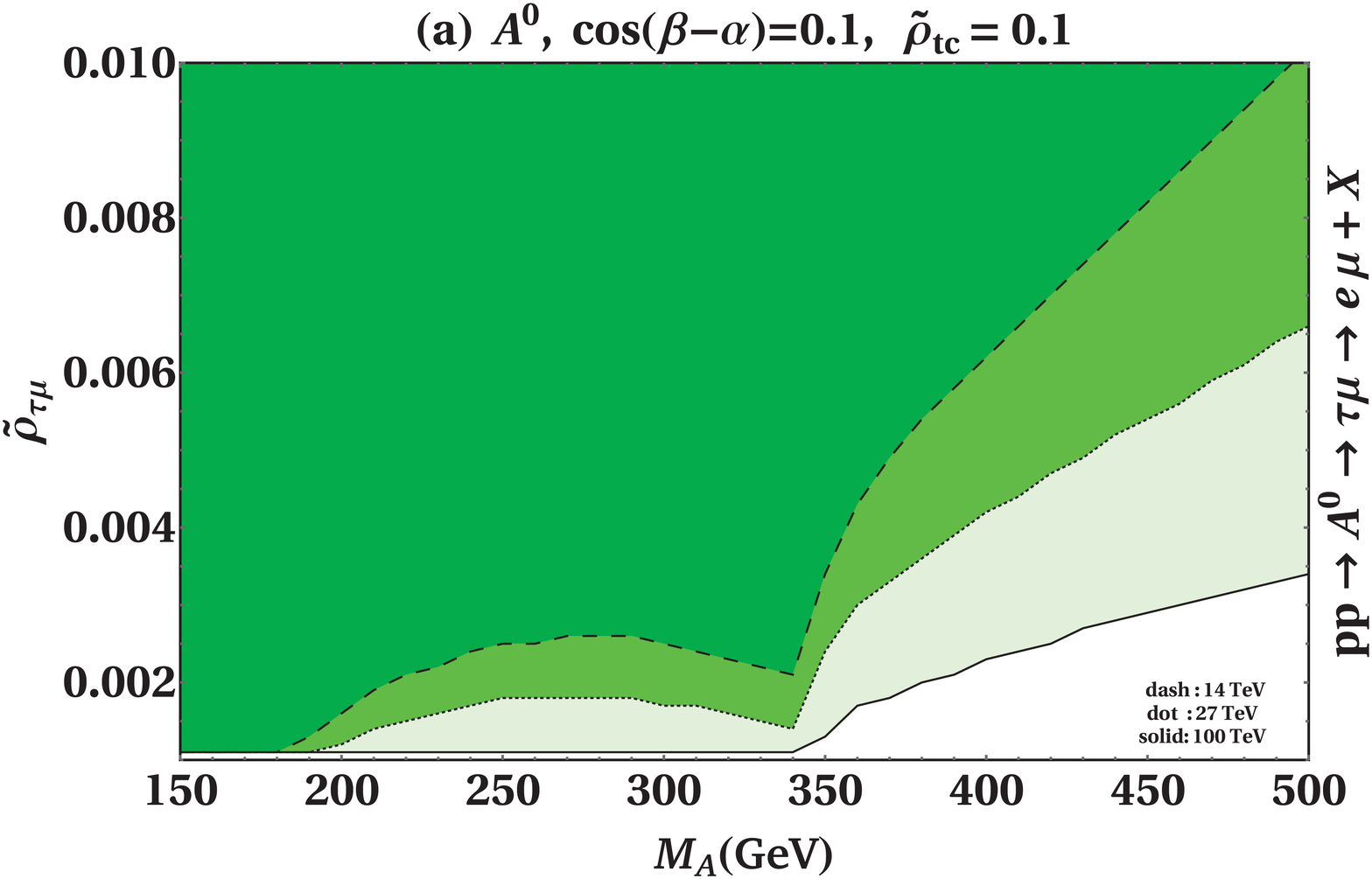}
\hspace{0.1in}
\includegraphics[width=78mm]{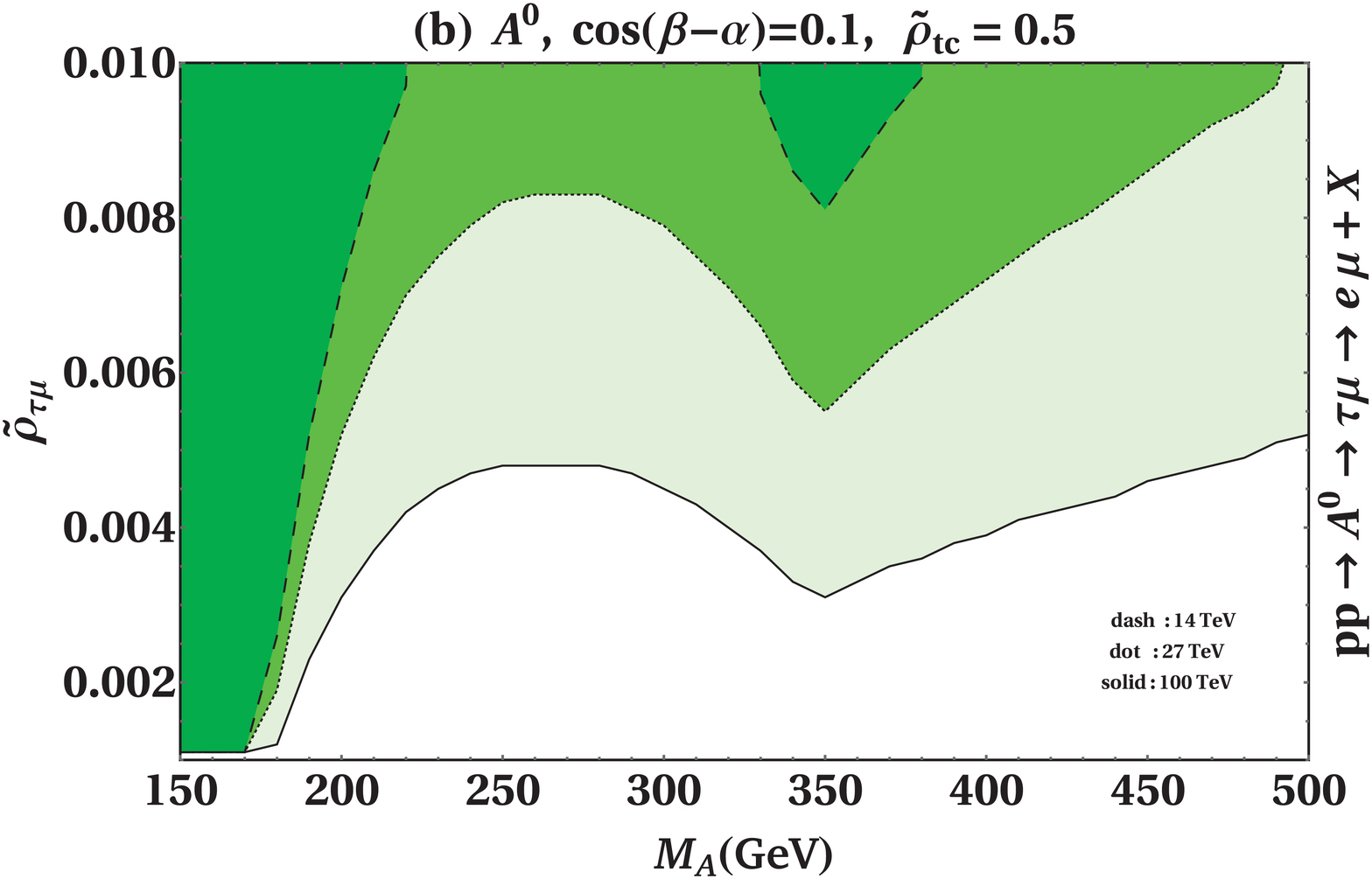} \\
\includegraphics[width=78mm]{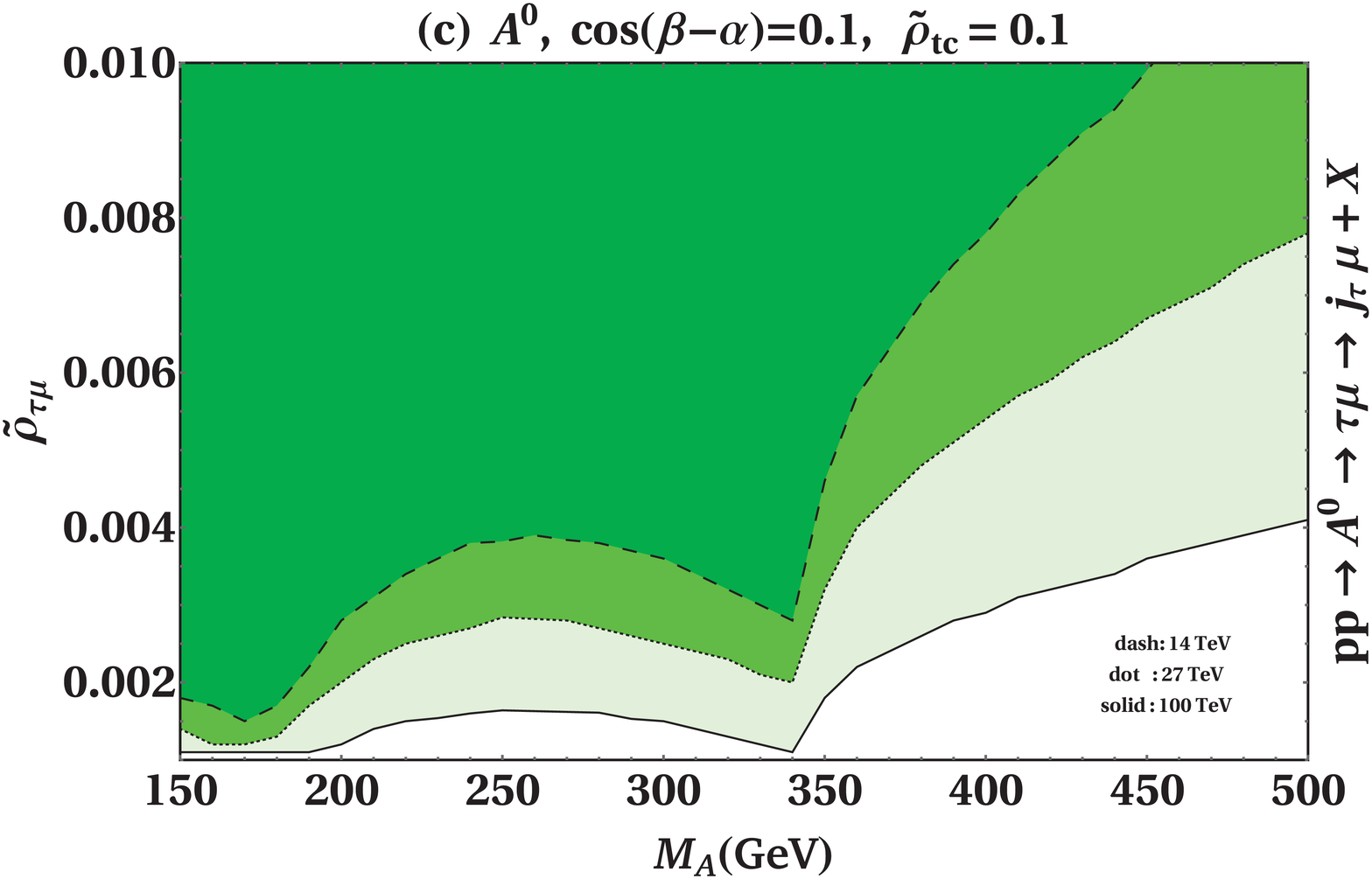}
\hspace{0.1in}
\includegraphics[width=78mm]{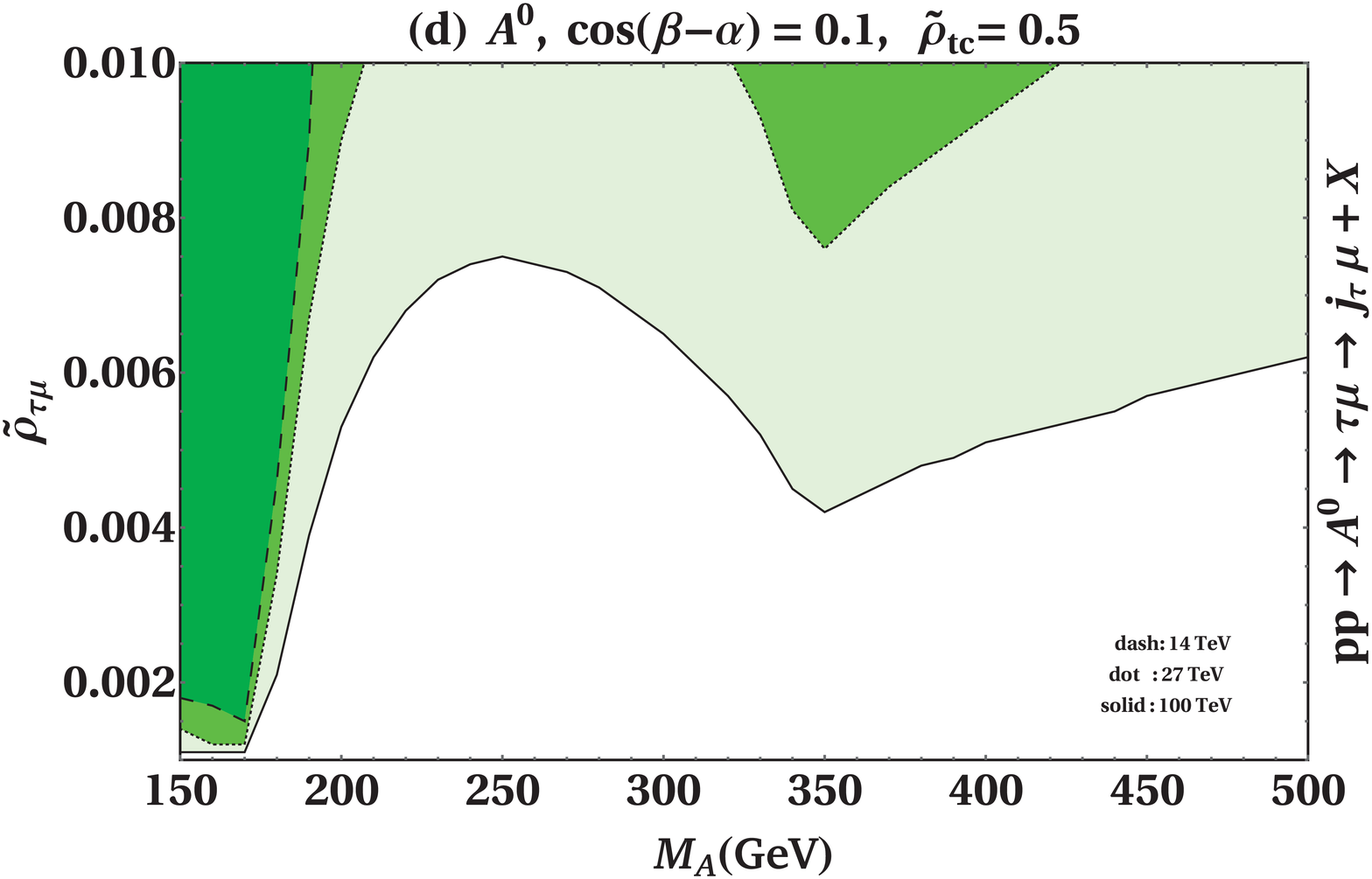} \\
\caption{
Discovery range at the LHC and future hadron colliders with
$\sqrt{s} = 14$ TeV (green dark shading),
27 TeV (intermediate shading) and 100 TeV (light shading)
for $pp \to A^0 \to \tau \mu +X$ in the ($M_A,\tilde\rho_{\tau\mu}$) plane.
We require $5\sigma$ significance for 3000 fb$^{-1}$. 
Top (bottom) row is for leptonic (hadronic) tau decay for
 $\tilde{\rho}_{tc}$ = 0.1 [(a) and (c)] and
 $\tilde{\rho}_{tc}$ = 0.5 [(b) and (d)].
}\label{fig:HAcontours}
\end{center}
\end{figure}

\subsection{Discovery Reach for Pseudoscalar $A^0$}

The pseudoscalar $A^0$ has higher production cross
section, and with no suppression coming from $A^0 \to h^0 h^0$, which is forbidden,
it is more promising than the heavy scalar $H^0$.
Fig.~\ref{fig:HAcontours} shows the discovery region for $pp \to A^0 \to \tau\mu +X$
in the ($M_A,\tilde\rho_{\tau \mu})$ plane, for $\tilde{\rho}_{tc}$ = 0.1 and 0.5,
including both the leptonic channel $\tau \to e\nu\nu$ (upper panels)
and the hadronic channel $\tau \to j_\tau \nu$ (lower panels).
Because of high QCD backgrounds,
performance for hadronic $\tau$ decay is worse than leptonic decay,
despite its higher branching ratios.

\begin{figure}[b]
\begin{center}

\includegraphics[width=78mm]{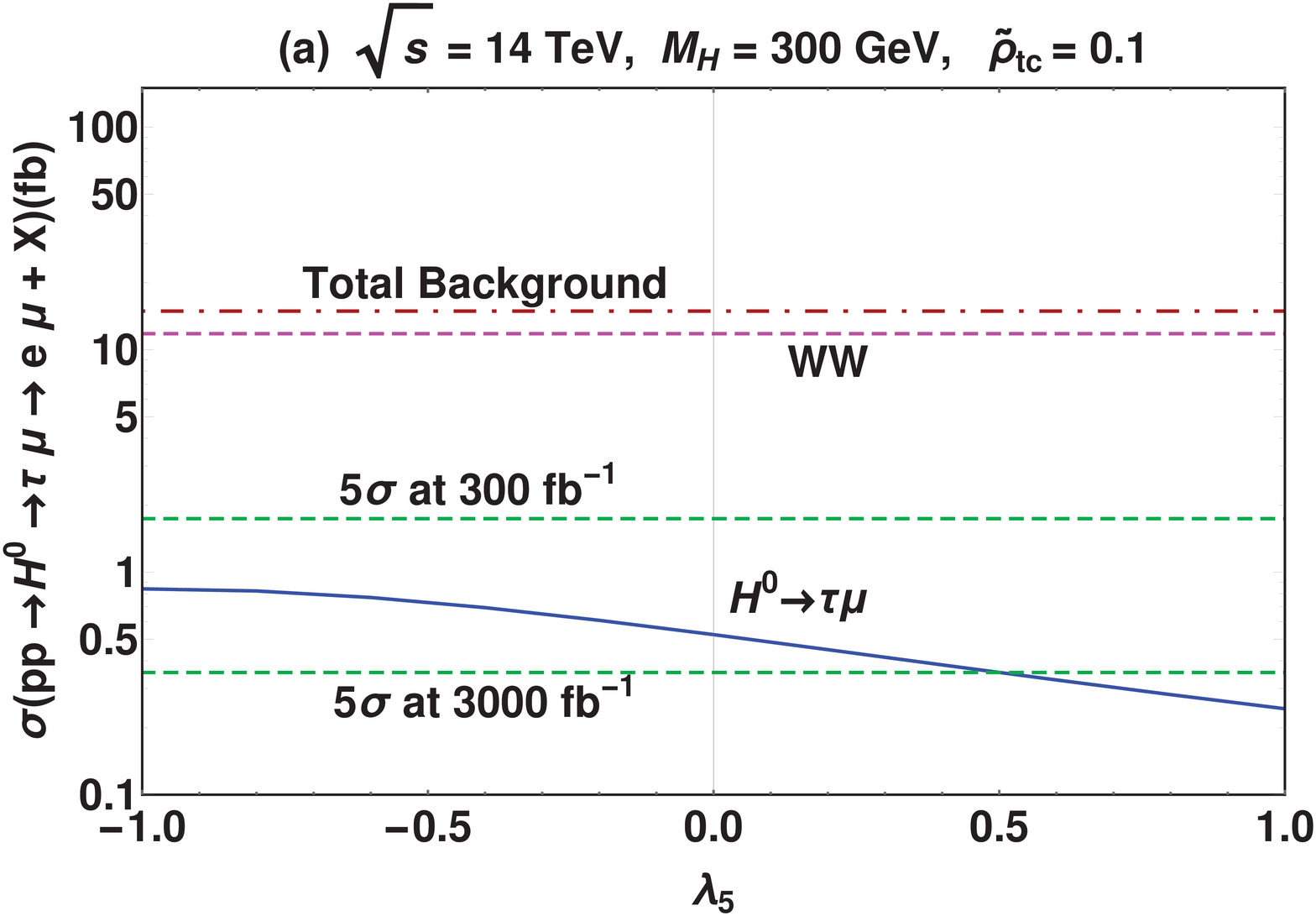}
\hspace{0.1in}
\includegraphics[width=78mm]{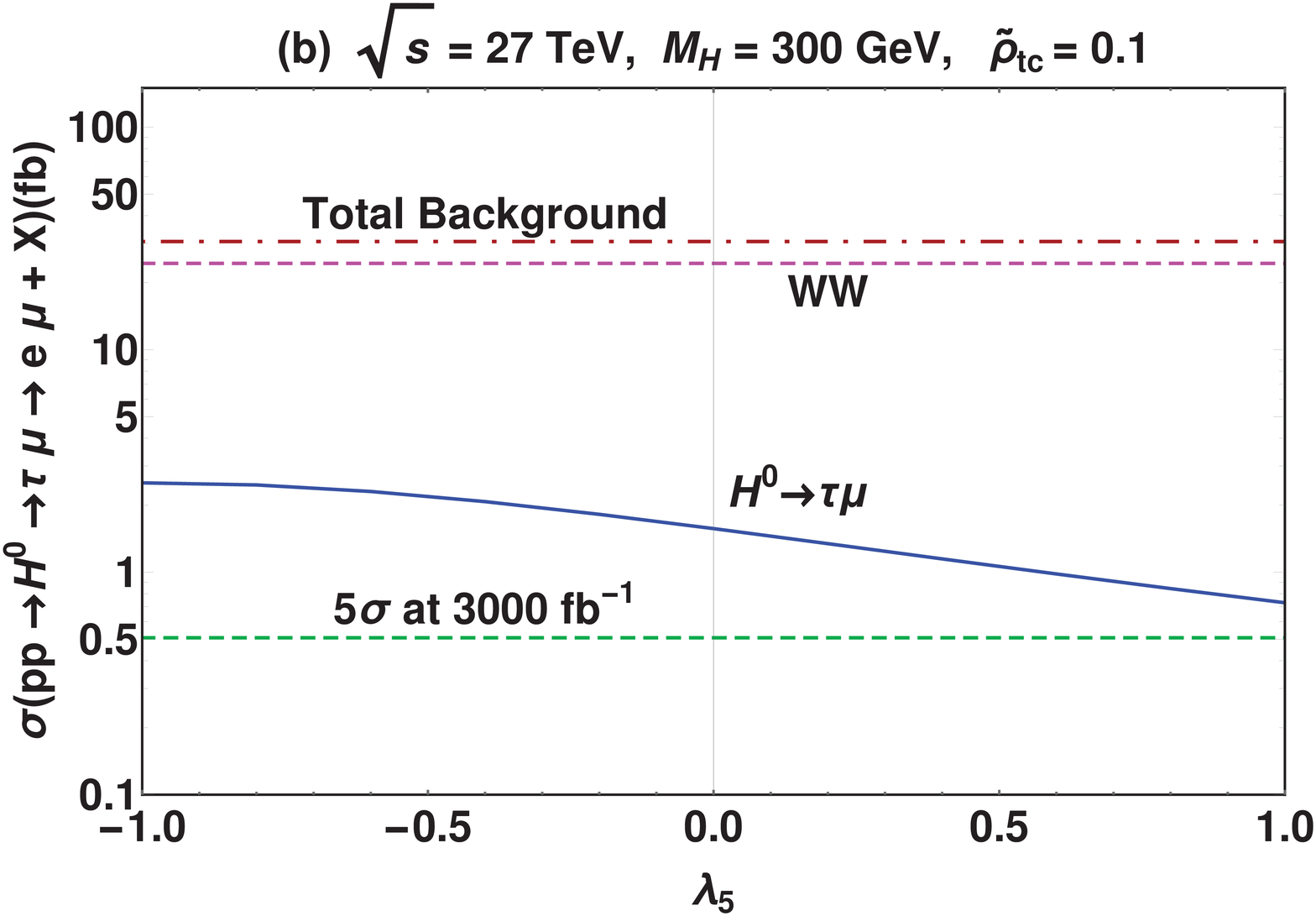} \\
\includegraphics[width=78mm]{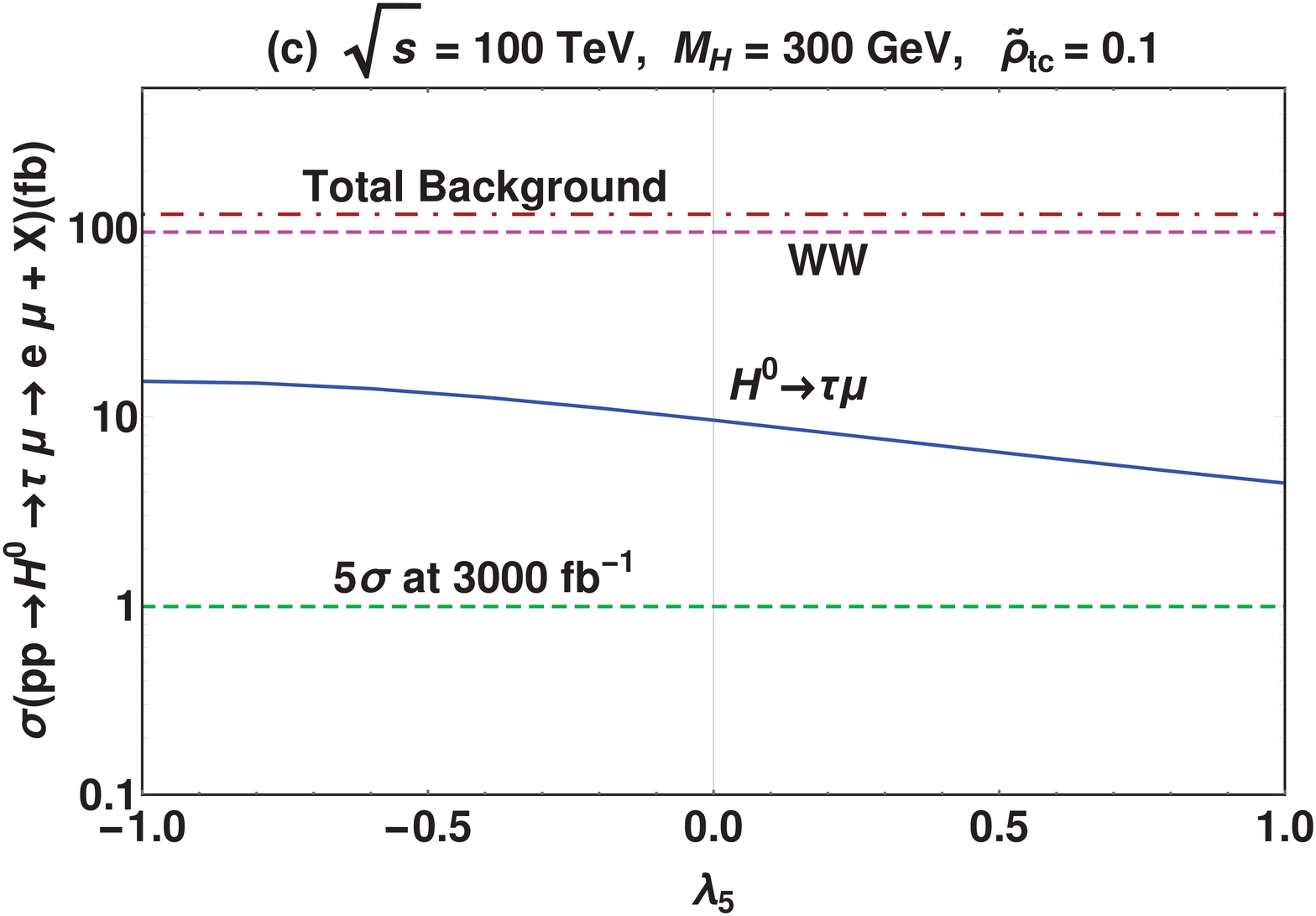}
\hspace{0.1in}
\includegraphics[width=78mm]{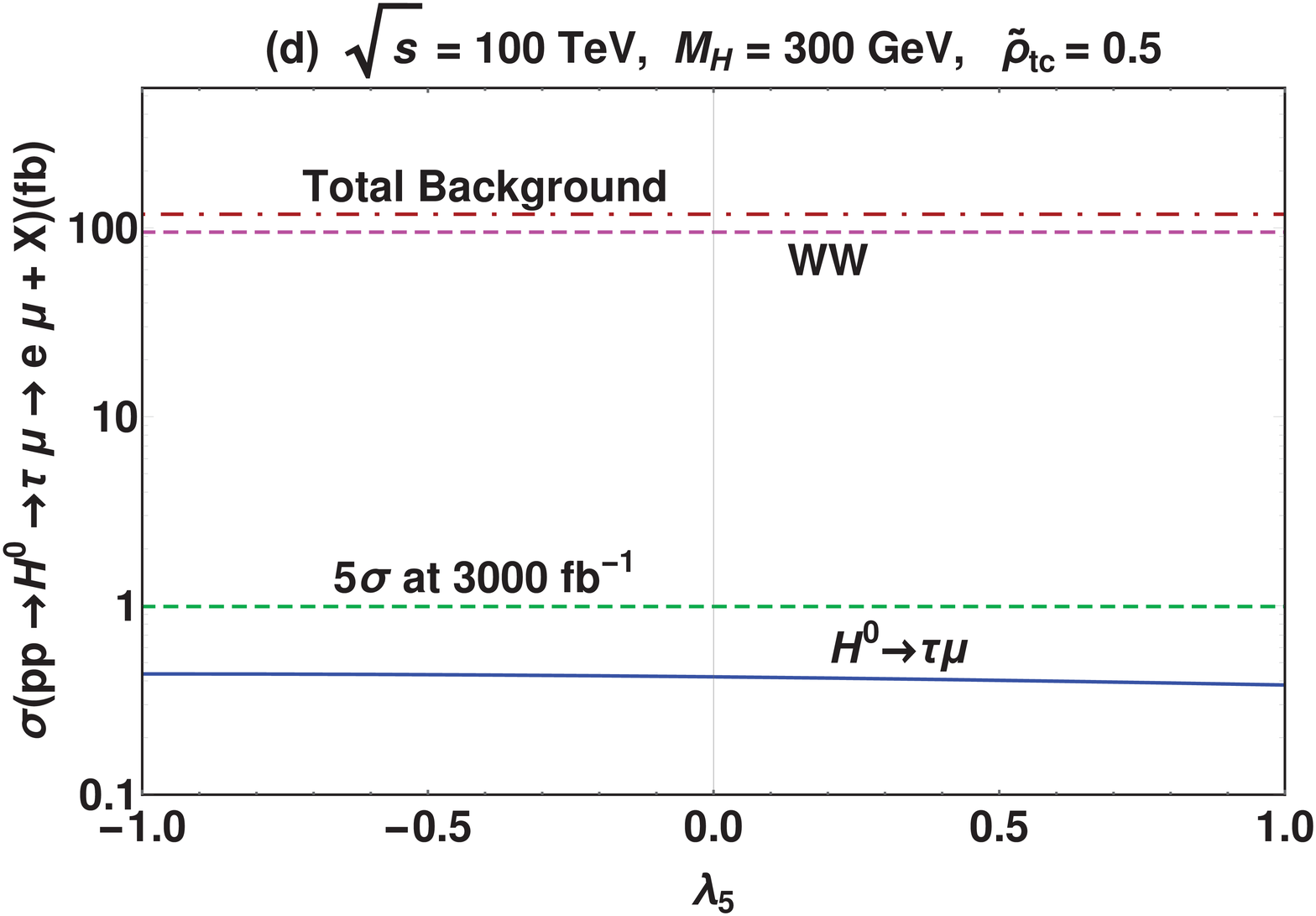} \\
\caption{
Cross section (in fb) of $pp \to H^0 \to \tau\mu \to e\mu + X$ (blue solid)
at (a) $\sqrt{s} = 14$ TeV, (b) 27 GeV, and (c) 100 TeV,
as a function of $\lambda_5$ with $M_H$ = 300 GeV,
$\rho_{\tau\mu} = 0.01$, $\tan\beta = 1$, $c_{\beta-\alpha} = 0.1$,
and $\tilde{\rho}_{tc} = 0.1$.
Also shown are the total background (maroon dotdash),
the predominant background from $pp \to W^+W^- +X$ (magenta dash),
and the 5$\sigma$ signal significance (green dash)
with integrated luminosity ${\cal L} = 3000$ fb$^{-1}$ 
or 300 fb$^{-1}$. 
We also present the results for
(d) $\tilde{\rho}_{tc} = 0.5$ at $\sqrt{s} = 100$ TeV.
}\label{fig:HHsigma}
\end{center}
\end{figure}

We show our results for $\sqrt{s}$ = 14, 27 and 100 TeV.
At low masses, $M_A < 180$ GeV or roughly the $t\bar c$ threshold,
the entire range of $\rho_{\tau \mu}$ is detectable
at 3000 fb$^{-1}$, independent of the center-of-mass energy.
For an intermediate range (200 GeV $< M_A <$ 300 GeV), our discovery
region starts shrinking because of $A^0 \to t\bar c$ predominance
(plus a milder effect from $A^0 \to Z h^0$ turn-on),
which is more striking for the larger $\tilde \rho_{tc} = 0.5$ value
as shown in the right panel plots of Fig.~\ref{fig:HAcontours}.
For higher mass range ($M_A > 300$ GeV),
we see a slight increase in the 5$\sigma$ region
before and around $M_A \sim 2 m_t$, owing to the rise in
production cross section for $g g \to A^0$,
before the turn-on of $A^0 \to t\bar{t}$ decay
further suppresses our signal towards higher masses
beyond $M_{A} \gtrsim 360$ GeV.
Note that $\tilde \rho_{tc} = 0.5$ is
actually larger than $\rho_{tt}$ for our mass range
 (see Eq.~(\ref{eq:rho_tt})),
which is constrained by B physics.

\subsection{Discovery Reach for Heavy CP-even Scalar $H^0$}

For the heavy CP-even boson $H^0$, the situation is quite different.
The branching fraction for $H^0 \to \tau \mu$ is affected by
$\rho_{tc}$, tan$\beta$ and $\lambda_5$.
The latter Higgs sector parameter affects the $H^0 \to h^0h^0$ decay,
where in Fig.~2 we illustrated with $\lambda_5 = 0$.
In order to understand the effect of $\lambda_{5}$, we perform a case
study for $pp \to H^0 \to \tau\mu \to e\mu +X$
with $M_H = 300$ GeV, $\rho_{\tau \mu} = 0.01$,
and scan over $-1 \leq \lambda_{5} \leq 1$ for tan${\beta}$ = 1.
The results are shown in Fig.~4 for $\sqrt{s} = 14$, 27
and 100 TeV for the leptonic channel and $\tilde \rho_{tc} = 0.1$.
The hadronic channel is similar except it will have higher
QCD background.

\begin{figure}[b]
\begin{center}

\includegraphics[width=78mm]{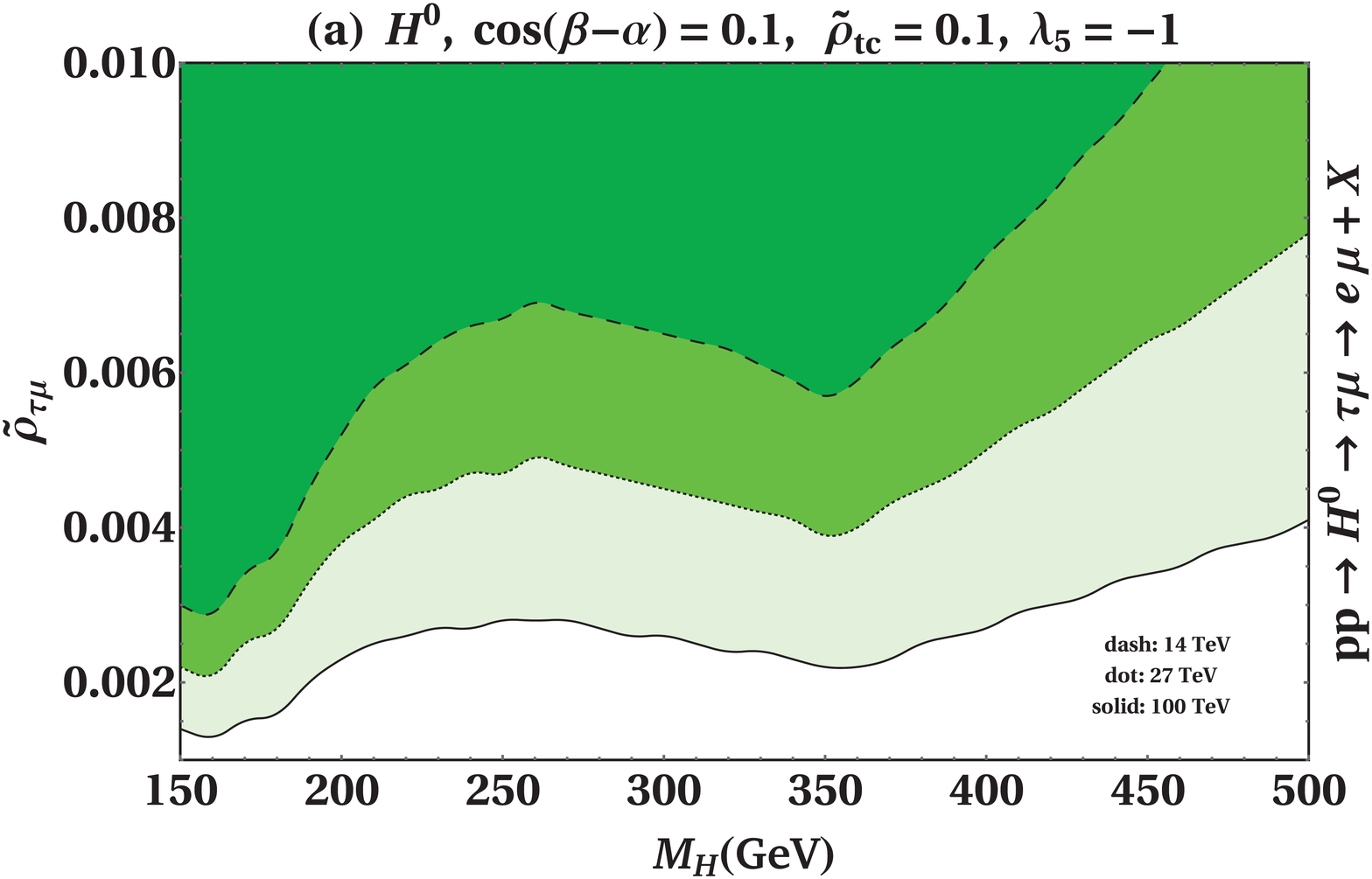}
\hspace{0.1in}
\includegraphics[width=78mm]{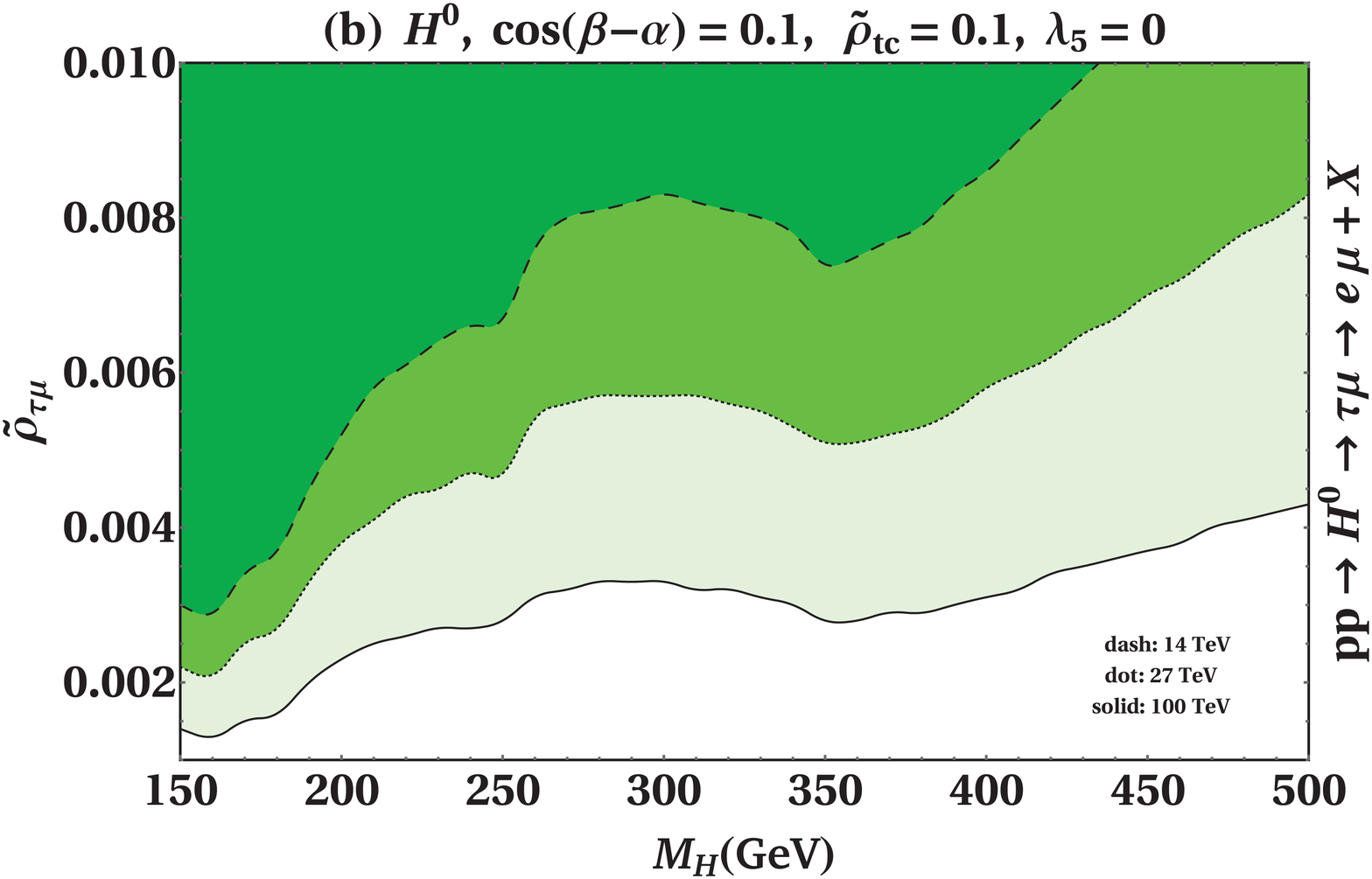} \\
\includegraphics[width=78mm]{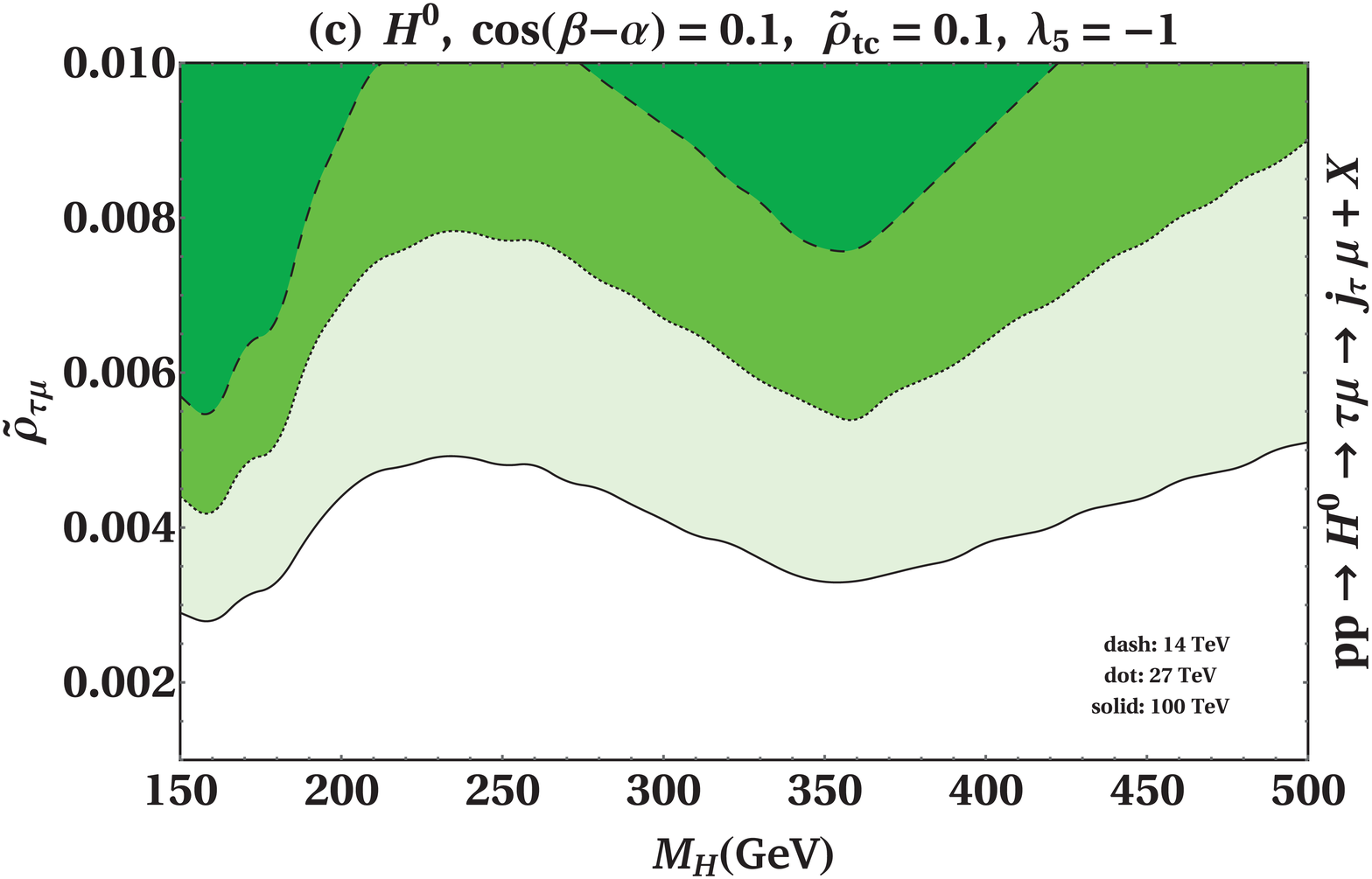}
\hspace{0.1in}
\includegraphics[width=78mm]{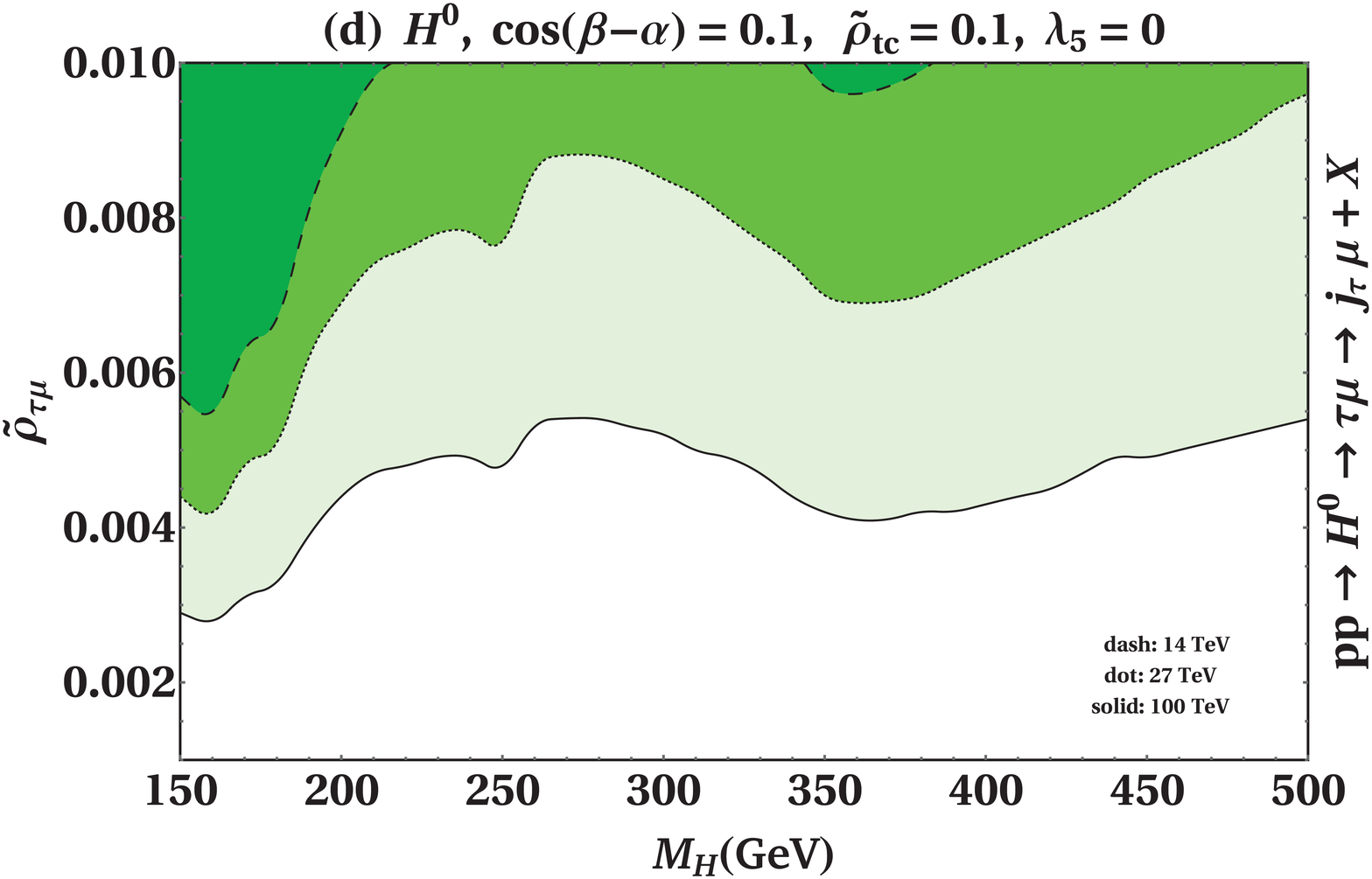}
\caption{
Discovery regions at the LHC and future hadron colliders with
the $\sqrt{s} = 14$ TeV (green dark shading),
27 TeV (intermediate shading) and 100 TeV (light shading)
for $pp \to H^0 \to \tau \mu +X$ in the ($M_H,\tilde\rho_{\tau\mu}$) plane.
We require at least $5\sigma$ significance for 3000 fb$^{-1}$.
Top (bottom) row is for leptonic (hadronic) tau decay
with
 $\lambda_5 = -1$ [(a) and (c)] and
 $\lambda_5 = 0$ [(b) and (d)].
} \label{fig:HHcontours}
\end{center}
\end{figure}

We observe that for a fixed value of $\tan\beta$,
increasing $\lambda_5$ from $-1$ to 0 lowers the cross section
of $pp \to H^0 \to \tau\mu +X$ while increasing the trilinear
Higgs coupling, $g_{Hhh}$, which enhances the branching fraction of
$H^0 \to h^0 h^0$,
\begin{eqnarray}
g_{Hhh} \simeq - \frac{c_{\beta-\alpha}}{v}
  \bigg[ 4 m_A^2 - 2m_h^2 - m_H^2  + 4 \lambda_5 v^2 \, ~ \bigg] \, ,
\end{eqnarray}
with $\lambda_6 = \lambda_7 = 0$.
As a case study, let us choose the values of $\lambda_5 = -1$, 0, 
with $\tan\beta = 1$ to preserve tree-level unitarity and
stability for a general 2HDM, which resembles the generic case more
closely, and perform a scan for $ 0.001 \leq \rho_{\tau \mu} \leq
0.01$ and 150 GeV $\leq M_H \leq$ 500 GeV.
The results are shown in Fig. 5.
There is a large discoverable region in the low mass regime
($M_H < 180$ GeV). However, as we start increasing $M_H$,
first $H^0 \to t \bar c$, then $H^0 \to h^0 h^0$, then $H^0 \to t\bar{t}$
become dominant.
The discovery potential is improved somewhat around
the $M_H \sim 2 m_t$ threshold because of rise in
$g g \to H^0$ production cross section,
appearing as `dips' of $5\sigma$ contours in Fig.~3 and Fig.~5.
Beyond that region, we still have some parameter space that can be probed,
and a 100 TeV high energy collider can probe to lower couplings.
The likelihood of detection increases as we reduce the value
of $\lambda_5$, from 0 to $-1$.
The situation for $\tilde \rho_{tc} \gtrsim 0.5$ would be worse than
$A^0 \to \tau\mu$, the right panels of Fig.~3.

\section{Conclusion}

The general two Higgs doublet model offers a very rich phenomenology
for flavor changing neutral Higgs interactions with fermions, because
of the absence of any symmetry to suppress them. Strong experimental
constraints exist for these FCNH interactions, but third generation
fermions might offer promising signatures for new physics at the LHC
and future hadron colliders.
Experimental data from LHC Run 1 had shown some hints
for the light CP-even Higgs boson $h^0 \to \tau \mu$,
but became insignificant with 2016 CMS data at Run 2.
However, in the general 2HDM, the coupling probed is
$\lambda_{h\tau\mu} = \rho_{\tau\mu}\cos(\beta-\alpha)$,
which is expected to be small in the alignment limit of
$\cos(\beta-\alpha) \to 0$, where the light CP-even Higgs
boson $h^0$ approaches the standard Higgs boson.

For heavy Higgs states, the pseudoscalar $A^0$ boson has
FCNH coupling $\lambda_{A\tau\mu} = \rho_{\tau\mu}$
that is independent of $\cos(\beta-\alpha)$, while
the heavy CP-even scalar $H^0$ has FCNH coupling
$\lambda_{H\tau\mu} = \rho_{\tau\mu}\sin(\beta-\alpha)$,
where $\sin(\beta -\alpha)$ is expected to be close to unity.
Thus, they offer great promise to discover FCNH signals
with lepton flavor violating production of
$pp \to H^0,A^0 \to \tau\mu +X$
at the LHC and future hadron colliders.

We have investigated the prospects of discovering
$H^0,A^0 \to \tau \mu$ for the high luminosity (HL)
and high energy (HE) LHC and future high energy $pp$ colliders.
With  gluon fusion  being the dominant mode of
production for both heavy scalars because of finite $\rho_{tt}$,
we find promising results for LHC with $\cos(\beta - \alpha) = 0.1$,
$\tilde{\rho}_{tc}  = 0.1$, when $H^0,A^0 \to t\bar{c}+\bar{t}c$ is
not yet overwhelming for $M_H$ up to 300 GeV.
The choice of $h^0$-$H^0$ mixing parameter $\cos(\beta - \alpha)$
and $\tilde\rho_{tc} = \sqrt{(|\rho_{tc}|^2 + |\rho_{ct}|^2)/2}$ values
are meant as illustrative.
Having taken degenerate $H^0$, $A^0$ and $H^+$,
$M_{H^0} < 300$ GeV can still evade $b\to s\gamma$ constraint
and should be taken seriously.
It should be noted that $A^0$ is more
promising than $H^0$ because of its higher production cross section
and fewer decay channels affecting its decay to $\tau \mu$,
but $H^0$ decay depends also on Higgs potential due to $h^0h^0$ mode.
If $\tilde{\rho}_{tc}$ is considerably larger than 0.1,
$H^0$, $A^0$ decay to $t\bar c$ would suppress $\tau\mu$ observability,
and a higher energy collider would be needed.

Our study has focused on discovering $H^0$, $A^0$ in $\tau\mu$ final state,
but companion final states such as
$Zh^0$ (for $A^0$), $h^0h^0$ (for $H^0$) and $t\bar c$, $t\bar t$
are worthy pursuits in their own right,
some of which have been studied elsewhere.

\medskip\medskip

\noindent{\bf Acknowledgments}.
MK thanks E. Senaha for useful discussion on the $\tau \to \mu \gamma$
constraint.
C.K. thanks the High Energy Physics Group at National Taiwan University
for hospitality, where part of the research was completed.
This research was supported in part by the U.S. Department of Energy
(RJ, CK, BM, and AS) as well as by grants MOST 106-2112-M-002-015-MY3,
107-2811-M-002-039, and 107-2811-M-002-3069 (WSH and MK).


\end{document}